\UseRawInputEncoding
\documentclass[
    reprint,
    doublecolumn,
    superscriptaddress,
    amsmath,amssymb,
    aps, prapplied,
showkeys
]{revtex4-2}
\usepackage{tikz}
\usepackage{makecell} 
\usepackage{multirow}
\usepackage{booktabs} 
\usepackage{bibentry}
\usepackage{float}
\usepackage{placeins}
\usepackage{fontspec}

\usepackage{graphicx}
\usepackage{dcolumn}
\usepackage{bm}
\usepackage{amsthm} 
\usepackage{braket}
\usepackage[colorlinks=true, allcolors=blue]{hyperref}
\usepackage{cleveref}
\usepackage{soul}
\usepackage{xcolor}

\usepackage{xcolor}
\usepackage{listings}
\usepackage{upquote}

\definecolor{codebg}{HTML}{F7F7F8}
\definecolor{codeframe}{HTML}{E2E2E6}
\definecolor{codecomment}{HTML}{6A737D}
\definecolor{codenumber}{HTML}{9AA0A6}

\lstdefinestyle{mos_shell}{
  language=bash,
  basicstyle=\CodeFont\small,      
  frame=single,
  rulecolor=\color{codeframe},
  frameround=tttt,
  framesep=6pt,
  xleftmargin=10pt,
  xrightmargin=6pt,
  breaklines=true,
  columns=fullflexible,
  keepspaces=true,
  showstringspaces=false,
  upquote=true,
  commentstyle=\color{codecomment}\itshape,
  numbersep=8pt,
  tabsize=2,
  breaklines=true,
  breakatwhitespace=false,
  prebreak=\mbox{\tiny$\hookleftarrow$}, 
}

\usepackage[ruled,vlined,linesnumbered]{algorithm2e}
\SetKwInput{KwInput}{Input}
\SetKwInput{KwOutput}{Output}
\SetKwRepeat{Do}{do}{while}

\definecolor{lightgreen}{RGB}{204,255,204}

\allowdisplaybreaks

\soulregister{\cite}{7}
\soulregister{\ref}{7}
\soulregister{\textbf}{7}

\sethlcolor{lightgreen}

\usepackage[normalem]{ulem} 
\usepackage{xcolor}
\newcommand{\delete}[1]{}

\newcommand{\noneed}[1]{}

\newcommand{\red}[1]{{}{#1}}
\newcommand{\redc}[1]{{}{#1}}

\newcommand{\redcnew}[1]{{}{#1}}
\makeatletter
\begin{document}

    \title{RASP: Reliability ab initio simulation package of MOSFETs based on all-state model}
    \author{Xinjing Guo}
    \affiliation{College of Integrated Circuits and Micro-Nano Electronics, and Key Laboratory of Computational Physical Sciences (MOE), Fudan University, Shanghai 200433, China}
    \author{Menglin Huang}
    \email{menglinhuang@fudan.edu.cn}
    \affiliation{College of Integrated Circuits and Micro-Nano Electronics, and Key Laboratory of Computational Physical Sciences (MOE), Fudan University, Shanghai 200433, China}
    \author{Shiyou Chen}
    \email{chensy@fudan.edu.cn}
    \affiliation{College of Integrated Circuits and Micro-Nano Electronics, and Key Laboratory of Computational Physical Sciences (MOE), Fudan University, Shanghai 200433, China}
    \affiliation{Shanghai Multi-scale Simulation Technology Co., Shanghai 201999, China}

    \begin{abstract}
	As transistors continue to scale down, device reliability has become a critical concern. In order to accurately simulate defect-induced reliability degradation in \redcnew{MOSFET based logic, memory and power} devices, we develop RASP (Reliability Ab initio Simulation Package), which implements the all-state model for reliability simulation. 
    Unlike conventional two-state and four-state models that consider only two and four defect configurations respectively, 
    the all-state model systematically considers all possible defect configurations in amorphous gate dielectrics and all nonradiative multiphonon (NMP) and thermal transition pathways among them. \redc{With defect parameters obtained from ab initio calculations as input}, RASP enables accurate simulation of threshold voltage shifts caused by defects. Using RASP to simulate oxygen vacancies in a-SiO$_\text{2}$​, we find that they are a non-negligible source of negative bias temperature instability (NBTI).
    
    \end{abstract}
    
    \maketitle

    \section{INTRODUCTION}
    \label{intro}

As field-effect transistor (FET) technology scales into the sub-10 nm regime, device reliability has become a critical concern that directly impacts integrated circuit performance and lifetime~\cite{Reisinger2014, RN48, RN49, RN51, RN70, RN71, RN72, mi15020269}. This challenge is common to all types of FETs, including planar metal-oxide-semiconductor FETs, nonplanar FinFETs, gate-all-around FETs (GAAFETs), and complementary FETs (CFETs)~\cite{RN48,RN49,RN51,RN70,RN71}. Among various reliability degradation mechanisms, negative bias temperature instability (NBTI) \cite{Grasser2013-kn, Mahapatra2021-pu, Mahapatra2015-jj}, random telegraph noise (RTN) \cite{5703436, PUGLISI2013160, 6963398, 6948806}, and time-dependent dielectric breakdown (TDDB) \cite{8844268, 8824213, Padovani2024-mf} have attracted significant attention. Despite their distinct macroscopic manifestations, these phenomena originate from the similar physical mechanism: defect generation and charge trapping/de-trapping processes within the gate dielectric and at its interfaces \cite{8844268, 6860643, ICIT_2014_Grasser, PhysRevApplied_YueYang_Liu, guo2024all}. Among these reliability issues, NBTI causes a negative shift in the threshold voltage ($V_{\mathrm{th}}$) and a degradation of the drive current in PMOS transistors. This effect becomes increasingly pronounced as device dimensions continue to shrink~\cite{Reisinger2014}, 
{\red{significantly slowing down transistor switching speed, increasing circuit delays and potentially resulting in timing failures,}}

which has attracted extensive attention from both academia and industry~\cite{Grasser2013-kn, Mahapatra2021-pu, Mahapatra2015-jj}. Therefore, a thorough understanding of the physical mechanisms underlying NBTI and the development of accurate models are crucial for circuit design and reliability assessment at advanced technology nodes~\cite{Grasser2013-kn}.

The physical understanding of NBTI has long revolved around two major physical pictures. The first is the classical reaction-diffusion (RD) model \cite{Mahapatra2018-de, Mahapatra2021-pu, Mahapatra2015-jj}. This model assumed that electric field and elevated temperature cause Si-H bonds at the Si/SiO$_\text{2}$ interface to break, generating interface defects ($N_\text{it}$) and mobile hydrogen species (such as H$^\text{+}$ or H$_\text{2}$). These hydrogen species subsequently diffuse away through the gate dielectric, causing long-term, permanent $V_\text{th}$ shift. The RD model successfully explains the power-law time dependence of NBTI degradation ($\Delta V_\text{th} \propto t^n$) \cite{Mahapatra2015-jj}, and was regarded as the standard model for many years \cite{RD1, RD2, RD3, RD4, RD5, RD6, RD7, RD8, RD9, RD10}. However, subsequent time-dependent defect spectroscopy (TDDS) experiments revealed significant and rapid recovery phenomena in NBTI degradation, which the classical RD model cannot quantitatively describe \cite{6529142, RN80, RN83}. To explain these complex behaviors of NBTI, the defect-centric picture has gradually become prevailing and is now widely accepted~\cite{NMP1,RN59,RN63,NMP4,NMP5,RN40}. This defect-centric picture attributes NBTI degradation primarily to holes in the channel tunneling into the gate dielectric and being captured by defects via nonradiative multiphonon (NMP) transitions \cite{RN3, RN34, RN40}. Within this physical picture, a series of models have been proposed, including: the two-state model (Kirton-Uren model \cite{RN85}, also known as the standard carrier trapping/de-trapping model \cite{RN86}), the three-state model (Harry-Diamond-Labs (HDL) switching trap model \cite{HDL}), and the four-state model \cite{RN54}. \red{These models are all proposed based on a limited number of $V_\text{O}$ configurations} and the associated carrier capture/emission pathways. For instance, the two-state model only considers two ground-state configurations: $V_\text{O}^0$ (dimer) and $V_\text{O}^+$ (back-projected) \cite{Switching_Grasser_2009}. The subsequent three-state and four-state models, \red{based on the bistability assumption~\cite{HDL, RN59}}, introduce metastable configurations at different charge states to explain various experimentally observed phenomena (such as switching traps and fixed traps \cite{RN69, RN64}). Specifically, the four-state model considers one ground state and one metastable configuration for neutral $V_\text{O}$, and one ground state and one metastable configuration for charged $V_\text{O}$, totaling four states. \delete{And the differences in local atomic environments caused by the structural disorder of amorphous materials are characterized by statistical distributions of defect parameters~\cite{RN59}.}\redcnew{Moreover, to account for defects in different local atomic environments caused by the structural disorder of amorphous materials, these models characterize such variations through statistical distributions of defect parameters~\cite{RN59}.} These defect-centric models have been extensively validated in academic research and have become the theoretical foundation for device reliability modeling and simulation \cite{NMP1, RN59, NMP4, NMP5, RN40}. Several reliability simulation package based on defect-centric models have been developed, such as Comphy \cite{Comphy3_Dominic_2023, Comphy_Rzepa_201849} and MARS \cite{MARS_Liu_2024}. Meanwhile, these models have also been adopted by the integrated circuit industry. TCAD tools such as Synopsys Sentaurus TCAD \cite{RN15}, Silvaco TCAD \cite{RN46}, and Global TCAD Solutions Minimos-NT \cite{GTS} have incorporated the two-state and four-state models. Furthermore, companies like IMEC and Infineon Technologies have integrated these models at the SPICE level to predict circuit aging effects \cite{iMEC_BTI, IMEC_2, IMEC_3, Infineon_Technologies}.

However, for defects in amorphous gate dielectrics (such as a-SiO$_{\mathrm{2}}$), \red{owing to} the long-range disorder and low symmetry of amorphous systems, extensive theoretical calculations and experimental characterizations have demonstrated that defects may exhibit much more complex structural configurations~\cite{Shluger2020, Strand_2018, XiangweiJiang2025, guo2024all, WILHELMER2022114801}\red{, therefore the defect configurations do not satisfy this simple bistability assumption}. \red{Taking V$_{\mathrm{O}}$ as an example, our previous work~\cite{guo2024all} showed that neutral V$_{\mathrm{O}}$ can have four types of configurations (Si-dimer, left-back-projected, right-back-projected, and double-back-projected), while +1 charged V$_{\mathrm{O}}$ can have seven types of configurations (the four mentioned above plus left-in-plane, right-in-plane, and twisted). Moreover, the ground-state configuration varies depending on the local atomic environment. This structural diversity arising from the disorder of the local amorphous network is beyond both the bistability assumption and the limited defect configurations considered in previous defect-centric models.}

Therefore, to fully capture the realistic physical behavior of defects in amorphous gate dielectrics and more accurately describe their impact on device reliability, we recently proposed the all-state model \cite{guo2024all}. Starting from the structural and energetic characteristics of defects in amorphous materials, this model systematically considers all possible defect configurations across different local atomic environments and charge states, as well as all carrier capture/emission and thermal transition pathways among them. Based on the all-state model, we found that the two-state and four-state models may overlook critical defect configurations and key pathways \cite{guo2024all}, potentially leading to misidentification of NBTI defect origins and consequently causing errors in predicting device reliability and long-term circuit aging effects. This raises an important question: how can the all-state model be utilized to rapidly and accurately predict the impact of gate dielectric defects on device reliability? The key to this question lies in: (i) developing methods for fast and accurate calculation of all carrier capture/emission rates and thermal transition rates in the all-state model under various device operating conditions; (ii) developing methods to calculate the time-dependent probabilities of defect in each state\redc{, i.e., defect charge states,} under the competition among all transition pathways in the all-state model; (iii) achieving accurate device-level simulation of threshold voltage shifts induced by defects in amorphous gate dielectrics.

To systematically address the above challenges, we develop a simulation package for transistor reliability: Reliability Ab initio Simulation Package (RASP), which enables efficient application of the more accurate all-state model to device-level reliability simulation. Specifically, through its modular design, RASP directly addresses the three key issues mentioned above: (i) RASP integrates the device model (see Sec.~\ref{sec:MOS}) and carrier capture/emission rate module (see Sec.~\ref{sec:capture_and_emission_rate}), enabling fast and accurate calculation of carrier capture and emission rates under various device operating conditions through high-performance parallel computing and Fourier transform methods. (ii) \redcnew{Using the transition rates, RASP can rapidly solve coupled differential equations to obtain the probabilities of defect in each state under time-varying device operating conditions (see Sec.~\ref{sec:all-state-model} and \ref{sec:master_eq}).} (iii) RASP couples microscopic defect behavior with macroscopic electrical characteristics, enabling dynamic and accurate simulation of threshold voltage shifts induced by a large number of defects (see Sec.~\ref{sec:master_eq}). This allows RASP to better simulate the impact of carrier capture/emission by defects in amorphous gate dielectrics on \redcnew{MOSFET based logic, memory and power devices}.
	
	\redc{\section{DEVICE ELECTROSTATICS MODEL}}
	\label{sec:MOS}
	\begin{figure}[htbp]
    \centering
    \includegraphics[width=0.49\textwidth]{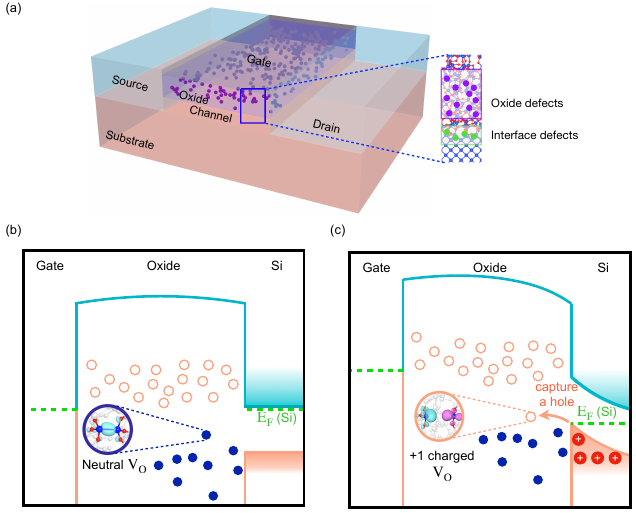}
    \caption{{Defects in the MOSFET gate dielectrics and the associated carrier capture process.} (a) Schematic of defects in the MOSFET gate dielectric. (b) Band diagram before and (c) after hole capture by an oxide defect.}
     \label{Fig:MOS-1}
    \end{figure}
    
	In metal-oxide-semiconductor field-effect transistors (MOSFETs), the gate oxide serves as the core insulating layer, and its integrity and electrical properties critically determine device performance~\cite{Taur_Ning_2021, Sze}. However, during fabrication or long-term operation, various defects are inevitably introduced within the gate oxide or at the oxide/semiconductor interface (as shown in Fig.~\ref{Fig:MOS-1}(a))~\cite{Kaczer2018GateOxideDefects, Cheung2022IntrinsicBreakdown, Zhang2022BTIReview, Process}. These defects, whether interface defects or bulk oxide defects, can capture carriers and form fixed or mobile charge centers in the gate oxide~\cite{Sze}. The presence of these charged defects significantly alters the internal electric field and electrostatic potential distribution (as shown in Fig.~\ref{Fig:MOS-1}(b-c)), thereby affecting key device parameters, particularly the threshold voltage ($V_{\mathrm{th}}$). Understanding and accurately simulating carrier capture/emission by defects in gate dielectrics and their impact on $V_{\mathrm{th}}$ is essential for improving device reliability and optimizing electrical characteristics. In the following, we will discuss in detail how charged defects affect transistor threshold voltage.

	\subsection{Gate bias equation of ideal MOS device}
	We first consider an ideal MOS structure without defects, where no charged defects exist in the oxide. For simplicity, we initially assume that the gate and semiconductor have the same work function, and later consider the case with a work function difference.

When the gate voltage is zero ($V_{\mathrm{G}}= 0$), no electric field exists within the MOS structure, and the semiconductor bands are in the flat-band condition with the surface Fermi level aligned to the bulk Fermi level~\cite{Sze}. When a non-zero gate voltage ($V_{\mathrm{G}}\neq 0$) is applied, surface charges are induced in the semiconductor, causing the Fermi level to bend near the surface~\cite{Sze}. The degree of band bending is characterized by the surface potential $\varphi_{\mathrm{S}}$, defined as the potential difference between the semiconductor surface and bulk. Meanwhile, the accumulated surface charges generate an electric field in the oxide, resulting in a voltage drop $V_{\mathrm{ox}}$ across the oxide. Thus, the gate voltage equation for an ideal MOS device without work function difference can be expressed as~\cite{Sze},

\begin{equation}
    \label{eq:1_simplified}V_{\mathrm{G}}= V_{\mathrm{ox}}+ \varphi_{\mathrm{S}}.
\end{equation}

Under the charge-sheet approximation \cite{BREWS1978345}, the voltage drop $V_{\mathrm{ox}}$ caused by the induced surface charge $Q_{\mathrm{S}}$ is given by,
\begin{equation*}
    V_{\mathrm{ox}}= Q_{\mathrm{S}}/ C_{\mathrm{ox}}
\end{equation*}
where $Q_{\mathrm{S}}$ includes all net charge contributions from the accumulation, depletion, and inversion layers, and is a function of the surface potential $\varphi_{\mathrm{S}}$. Therefore, the gate voltage equation becomes,
\begin{equation}
    \label{eq:2_simplified}V_{\mathrm{G}}= \frac{Q_{\mathrm{S}}(\varphi_{\mathrm{S}})}{C_{\mathrm{ox}}}+ \varphi_{\mathrm{S}}.
\end{equation}

In practical MOS structures, a work function difference ($\Phi_{\mathrm{MS}}$) typically exists between the gate and the semiconductor, defined as the difference between the gate work function $E_{\mathrm{W,gate}}$ and the semiconductor work function $E_{\mathrm{W,chan}}$. To compensate for this difference, a flat-band voltage ($V_{\mathrm{FB}}$) must be applied to align the gate and semiconductor Fermi levels~\cite{Sze},
\begin{equation*}
    \label{eq:VFB}V_{\mathrm{FB}}= \Phi_{\mathrm{MS}}= \frac{E_{\mathrm{W,gate}}- E_{\mathrm{W,chan}}}{q},
\end{equation*}
where $q$ is the elementary charge. Accounting for the work function difference, the complete gate voltage equation for an ideal MOS device becomes,
\begin{equation}
    \label{eq:2_revised}V_{\mathrm{G}}= \frac{E_{\mathrm{W,gate}}- E_{\mathrm{W,chan}}}{q}+ \frac{Q_{\mathrm{S}}(\varphi_{\mathrm{S}})}{C_{\mathrm{ox}}}+ \varphi_{\mathrm{S}}.
\end{equation}

This equation describes the response of the semiconductor surface potential to the gate voltage, and serves as the foundation for analyzing MOS device behavior under different bias conditions.

	\subsection{Gate bias equation of MOS device \redc{with} defects}
	
	In practical MOS structures, as shown in Fig.~\ref{Fig:MOS-1}(a), various charged defects often exist in the gate oxide. These defects may originate from impurity incorporation during fabrication, radiation damage, or thermal stress~\cite{Process}. Charged defects include fixed charges ($Q_\text{f}$), interface defects ($Q_{\mathrm{it}}$), and bulk oxide defects ($Q_{\mathrm{ox}}$). According to Poisson's equation $\nabla^{2}\phi = -\rho / \epsilon$, any charged defect in the oxide leads to a non-uniform potential distribution. Assuming an oxide thickness of $t_{\mathrm{ox}}$ and a defect charge density $\rho(x)$, where $x$ ranges from the channel/oxide interface ($x=0$) to the gate ($x=t_\mathrm{ox}$), the potential $\phi(x)$ in the oxide satisfies,
\begin{equation}\label{eq:possion}
    \frac{d^{2}\phi}{dx^{2}}= -\frac{\rho(x)}{\epsilon_{\mathrm{ox}}}.
\end{equation}

By integrating twice and applying boundary conditions, the contribution of charged defects to the total voltage drop across the gate oxide, $V_{\mathrm{traps}}$, is given by,
\begin{equation*}
\label{eq:6}
\begin{aligned}
V_{\mathrm{traps}}
&= -\frac{t_{ox}}{\epsilon_{\mathrm{ox}}}\int_{0}^{t_{\mathrm{ox}}}\rho(x)\left(1-\frac{x}{t_{\mathrm{ox}}}\right)\,dx \\
&= -\frac{1}{C_{\mathrm{ox}}}\int_{0}^{t_{\mathrm{ox}}}\frac{\rho(x)\left(t_{\mathrm{ox}}-x\right)}{t_{\mathrm{ox}}}\,dx .
\end{aligned}
\end{equation*}

For interface defects, the charge distribution can be treated as a Dirac delta function $\rho(x) = Q_{\mathrm{it}}\delta(x)$, and the integral simplifies to,
\begin{equation*}
    \label{eq:5}V_{\mathrm{traps}}= -\frac{Q_{\mathrm{it}}}{C_{\mathrm{ox}}},
\end{equation*}
where $Q_{\mathrm{it}}$ is the interface defect charge density.

For defects within the gate dielectric, assuming the charge distribution satisfies the charge-sheet approximation \cite{BREWS1978345}, the contribution $V_{\mathrm{traps}}$ can be expressed as,
\begin{equation}
    \label{eq:6_discrete}V_{\mathrm{traps}}= -\sum_{n=1}^{N}\frac{q_{\mathrm{T},n}}{C_{\mathrm{ox}}} \left(1 - \frac{x_{\mathrm{T},n}}{t_{\mathrm{ox}}}\right),
\end{equation}
where $N$ is the total number of charged defects, $q_{\mathrm{T},n}$ is the charge of the $n$-th defect given by $q_{\mathrm{T},n}= q \cdot \mathrm{P}_{\mathrm{T},n}$ with $\mathrm{P}_{\mathrm{T},n}$ being the probability of the defect in charged state, and $x_{\mathrm{T},n}$ is its position relative to the channel/oxide interface.

Therefore, the gate voltage equation for an MOS structure with charged defects becomes,
\begin{equation}
    \label{eq:4}V_{\mathrm{G}}= \frac{E_{\mathrm{W,gate}}- E_{\mathrm{W,chan}}}{q} + \frac{Q_{\mathrm{S}}(\varphi_{\mathrm{S}})}{C_{\mathrm{ox}}}+ \varphi_{\mathrm{S}} + V_{\mathrm{traps}}.
\end{equation}
	
	\redc{\subsection{Impact of charge trapping defects on the V$_\text{th}$ of MOS device}}
	By comparing Eq.~\eqref{eq:2_revised} and \eqref{eq:4}, it is evident that the threshold voltage of an MOS device with charged defects in the gate oxide shifts by $V_{\mathrm{traps}}$ relative to an ideal defect-free MOS device. This shift arises from the change of the oxide electrostatic potential distribution induced by charged defects, which alters the effective gate control. Among all charged defects, interface defects and bulk oxide defects can dynamically exchange carriers with the channel or gate (as shown in Fig.~\ref{Fig:MOS-1}(c)). This carrier exchange process changes the occupation probability of these defect centers in real time, causing dynamic variations in $V_{\mathrm{traps}}$ and ultimately leading to real-time threshold voltage shift. Interestingly, as discussed in Sec.~\ref{sec:capture_and_emission_rate}, carrier capture and emission rates are influenced by the internal electrostatic potential distribution, which, according to Eq.~\eqref{eq:4}, directly depends on $V_{\mathrm{traps}}$. Therefore, to accurately describe this dynamic process, the coupling between carrier capture/emission and the electrostatic potential distribution in the gate oxide must be considered, and the threshold voltage shift should be simulated through self-consistent solution of Eq.~\eqref{eq:4}.

    \ 
    
	\redc{\section{TRANSITION RATE MODEL}}\label{sec:capture_and_emission_rate}
	As discussed in Sec.~\ref{sec:MOS}, $\Delta V_\mathrm{{th}}$ is directly related to carrier capture and emission at oxide defects. Therefore, accurate calculation of carrier capture and emission rates is essential for quantitatively evaluating $\Delta V_\mathrm{{th}}$. Carrier capture/emission by defects in the gate dielectric inevitably involves tunneling between the defect and the channel, which can be described using a one-dimensional quantum tunneling model (see Sec.~\ref{sec:WKB}). For the carrier capture process (the emission process proceeds in the reverse direction), after a carrier tunnels to the vicinity of a defect, it has a certain probability of being captured. This involves two types of transitions: NMP transitions (see Sec.~\ref{sec:NMP}) and thermal transitions (see Sec.~\ref{sec:Thermal}). This section will discuss in detail how to calculate the transition rates for these three processes.

	\subsection{Carrier tunneling rate}\label{sec:WKB}
	Within the Wentzel–Kramers–Brillouin (WKB) approximation~\cite{wentzel1926verallgemeinerung,kramers1926wellenmechanik,Brillouin:1926blg}, the carrier tunneling probability is determined by the defect position ($x_\text{T}$), the position-dependent conduction band minimum (CBM) or valance band maximum (VBM) level of oxide ($E_\text{ox}^\text{VBM/CBM}(x)$), and the carrier energy ($E$),

	\begin{equation}\label{WKB}
	\begin{aligned}
	&f_\text{WKB}\!\left(E_\text{ox}^\text{VBM/CBM}(x),E\right)\\
	&=
	\exp\Biggl[-2\int_{0}^{x_\text{T}}
	\sqrt{\frac{2m_{t}}{\hbar^2}\Bigl(E_\text{ox}^\text{VBM/CBM}(x)-E\Bigr)}\,dx\Biggr].
	\end{aligned}
	\end{equation}
	where $m_t$ is the tunneling effective mass.
	
	\redc{\subsection{NMP transition rate}}\label{sec:NMP}
	\subsubsection{Detailed balance}\label{sec:detailedbalance}
	
	At equilibrium, defect occupation does not change with time, so carrier capture and emission must satisfy detailed balance, i.e., the capture rate into an occupied state equals the emission rate out of that state,
	\begin{equation}
    k_{c}f=k_{e}(1-f)
\end{equation}
where $f$ is the Fermi-Dirac occupation at the defect level. Substituting the explicit form of $f$, we obtain,
\begin{equation}
    \label{eq:H-eq2}k_{e}=k_{c}\exp \left(-\frac{E_{T}-E_{F}}{k_{B}T}\right),
\end{equation}

In semiconductor physics, the capture rate is typically expressed as the product of carrier concentration and capture coefficient, i.e., $k_{c}=C_{p}p$~\cite{RN26}. Substituting into Eq.~\eqref{eq:H-eq2} yields,
\begin{equation}
\label{eq:H-eq3}
\begin{aligned}
k_{e}
&= C_{p}N_{v}\exp\left(-\frac{E_{F}-E_{v}}{k_{B}T}\right)\cdot
   \exp\left(-\frac{E_{T}-E_{F}}{k_{B}T}\right) \\
&= C_{p}N_{v}\exp\left(-\frac{E_{T}-E_{v}}{k_{B}T}\right)
\end{aligned}
\end{equation}

From Eq.~\eqref{eq:H-eq3}, we see that $k_{e}$ is independent of carrier concentration and depends only on two defect characteristics: the carrier capture coefficient $C_{p}$ and the defect level $E_{T}$. Therefore, accurate calculation of the capture coefficient is the key to determining $k_{c}$ and $k_{e}$.
	\subsubsection{Fermi's golden rule}
	
\begin{figure}[htbp!]
        \centering
        \includegraphics[width=0.42\textwidth]{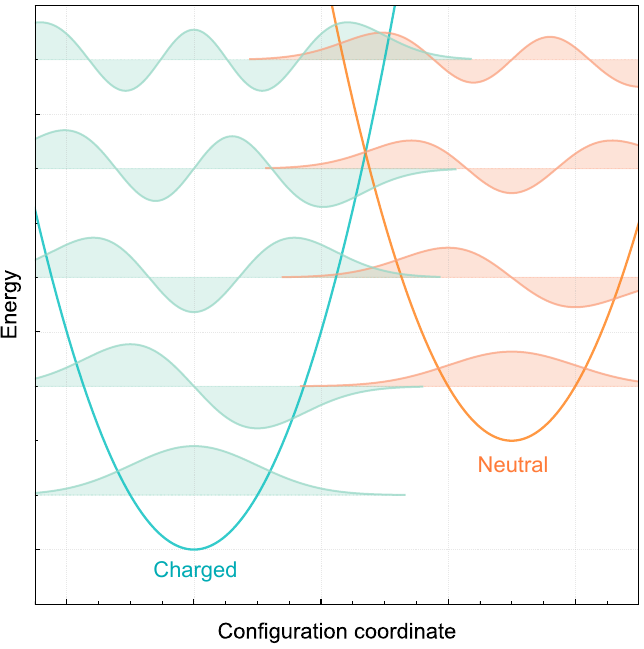}
        \caption{\red{Configuration coordinate diagram and phonon wavefunction for nonradiative multiphonon transitions.} }
        \label{Fig:H-1}
    \end{figure}

	In solids, the dominant mechanism for carrier capture is phonon-assisted electronic transitions. This theory was originally proposed by Huang and Rhys in 1950, who termed it nonradiative multiphonon (NMP) transitions \cite{RN65, Huang01111981}. Over the past decade, it has been successfully applied in first-principles calculations \cite{RN26}. The basic physical picture can be understood through the configuration coordinate diagram shown in Fig.~\ref{Fig:H-1}. Carrier capture/emission is accompanied by a change in the defect’s equilibrium configuration, i.e., a displacement of the minima of the two potential energy surfaces along $Q$. As a result, the vibrational wavefunctions of the initial and final states are not strictly orthogonal, and their overlap integral becomes non-zero. During this process, the system can absorb or emit an arbitrary number of phonons to satisfy energy conservation.

Although the physical picture of multiphonon transitions is clear, calculating the transition rate is complex. According to Fermi's golden rule,
\begin{equation}\label{eq:H-4}
    \begin{aligned}
    	r&=\\
    	&\frac{2 \pi}{\hbar}\sum_{m,n}\rho_{m}\left|\left\langle \Phi_{i m}(x, Q)|H_{e L}| \Phi_{f n}(x, Q)\right\rangle\right|^{2}\delta \left(E_{i m}-E_{f n}\right),
    \end{aligned}
\end{equation}
where $\Phi_{im}(x,Q)$ is the total wavefunction of the system in the initial state (including both lattice and electronic parts), $H_{eL}$ is the electron-phonon coupling matrix element, and $\rho_{m}$ is the occupation probability of the initial state. In practice, many studies adopt the Condon approximation for nonradiative transitions, which assumes the electron-phonon coupling matrix element to be constant and independent of the lattice coordinate $Q$~\cite{RN65}. However, Huang and Gutsche pointed out that the Condon approximation introduces significant errors in nonradiative transition calculations, and non-Condon effects must be considered~\cite{Huang01111981, Gutsche1982}. When the electron-phonon coupling varies linearly with $Q$, the coupling term takes the form,
\begin{equation}
    \label{eq:H-5}
    \begin{aligned}
    	&\left\langle\Phi_{i m}(x, Q)\right| H_{e L}\left|\Phi_{f n}(x, Q)\right\rangle\\
    	&=\left\langle\psi_{i}(x)\right| \frac{\partial H_{e L}}{\partial Q_{k}}\left|\psi_{f}(x)\right\rangle\left\langle\chi_{m}(Q)\right| Q_{k}\left|\chi_{n}(Q)\right\rangle,
    \end{aligned}
\end{equation}
where the first term is the derivative of the electronic wavefunction overlap with respect to $Q_k$, and the second term is the phonon wavefunction overlap containing the linear $Q_k$ term. Both terms correspond to the $k$-th phonon mode, and the system has $3N$ phonon modes in total. If the defect equilibrium structure remains unchanged upon carrier capture, $\chi_{m}$ and $\chi_{n}$ are orthogonal, and the matrix element is strictly zero. If lattice relaxation breaks this orthogonality, all phonon modes must be rigorously considered following the above equation; see Ref.~\cite{PhysRevB.111.115202} for details.

However, this approach is difficult to implement for carrier capture by defects in MOSFETs. Defects in devices exhibit spatial and energy distributions, and under bias, each defect has distinct properties (such as distance from the semiconductor interface and defect level relative to the valence band maximum). This requires independent transition rate calculations for each defect, making full consideration of all phonon modes computationally prohibitive.

	\subsubsection{Single-mode approximation}
	Over the past few decades, a widely adopted approach in solid-state physics and quantum chemistry for treating multiphonon problems is to approximate the contributions of all $3N$ phonon modes to electronic transitions by a single effective phonon mode~\cite{RN65,RN26}. This approach significantly reduces computational cost while maintaining reasonable agreement with full phonon calculations~\cite{ShiPRB2015, WickPRB2018}. In 2012, Alkauskas et al. defined such a special vibrational mode based on DFT calculations, i.e., the phonon mode that contributes most to electronic transitions is assumed to have a vibrational direction identical to the defect lattice relaxation direction~\cite{RN25}. Although this special mode is not an eigenmode of the real lattice, it effectively represents all eigenmodes. Based on this definition, the lattice relaxation along this direction can be expressed as,
\begin{equation}
    \label{eq:H-6}\Delta Q=\sqrt{\sum_{\alpha}m_{\alpha}\left(R_{f}-R_{i}\right)^{2}},
\end{equation}

By linearly interpolating between the two defect equilibrium structures according to Eq.~\eqref{eq:H-6}, the configuration coordinate diagram can be directly constructed from single-point energy calculations. Under the single-mode approximation, the matrix element in Eq.~\eqref{eq:H-5} involves only one phonon mode, and Eq.~\eqref{eq:H-4} simplifies to,
\begin{equation}
    \label{eq:H-3p}r=\frac{2 \pi}{\hbar}\left|W_{i f}\right|^{2}\sum_{m}\sum_{n} \rho_{m}\left|\left\langle\chi_{m}|Q| \chi_{n}\right\rangle\right|^{2}\delta \left(E_{i m}-E_{f n}\right),
\end{equation}

The transition rate calculation thus separates into two parts: the phonon overlap term $\left\langle\chi_{m}\middle| Q\middle|\chi_{n}\right\rangle$ and the electronic overlap term $W_{if}$. In the following, we discuss the calculation of each term and their implementation in RASP.

	\subsubsection{Phonon overlap intergral}
	The second part of Eq.~\eqref{eq:H-3p} is called the lineshape function, which can be written as,
\begin{equation}
    \label{eq:H-7}G=\sum_{m}\sum_{n}\rho_{m}\left|\left\langle\chi_{m}\left|Q\right|\chi_{n}\right\rangle\right|^{2}\delta\left(E_{i m}-E_{f n}\right).
\end{equation}

Over the past few decades, the calculation of phonon wavefunction overlap integrals under the single-mode approximation has been well established. The main approaches are based on the harmonic approximation to represent the wavefunctions of one-dimensional harmonic oscillators, including the polynomial method~\cite{Turianskynonrad2021}, the WKB approximation method~\cite{Jakob2021TED}, and direct solution of the Schrödinger equation~\cite{Kim2019PRB}. However, we note that all three methods perform calculations in energy space. According to Eq.~\eqref{eq:H-7}, to ensure convergence of the lineshape function $G$, overlap integrals must be computed for $m \times n$ vibrational states. For example, if $m=n=100$ and the device contains 10,000 defects, $10^{8}$ overlap integrals would be required at each time step of a transient simulation, which is clearly inefficient.

We used the Fourier transform method to improve the efficiency. This is performed by a single integration from $-\infty$ to $+\infty$ over time, which can replace the original $m \times n$ calculations, greatly improving simulation efficiency. We build upon the generating function method proposed by Lax~\cite{Lax} and Kubo~\cite{Kubo} under the Condon approximation, and further incorporate non-Condon effects (detailed derivation in Ref. \cite{PhysRevB.111.115202}). The approach is as follows:

We first define two time-dependent variables,
\begin{equation}
    \label{eq:H-8}\tau_{i}=-t-i \beta, \quad \tau_{f}=t
\end{equation}

Under the single-mode approximation, the phonon frequency before transition is $\omega_{i}$ and after transition is $\omega_{f}$. We then define several frequency-related quantities,
\begin{equation}
    \label{eq:H-9}a_{i}=\frac{\omega_{i}}{\sinh \left(\tau_{i}\hbar \omega_{i}\right)}, \quad a_{f}=\frac{\omega_{f}}{\sinh \left(\tau_{f}\hbar \omega_{f}\right)},
\end{equation}
\begin{equation}
    \label{eq:H-10}c_{i}=\frac{\omega_{i}}{\hbar}\operatorname{coth}\left(\frac{\tau_{i}\hbar \omega_{i}}{2}\right), \quad c_{f}=\frac{\omega_{f}}{\hbar}\operatorname{coth}\left(\frac{\tau_{f}\hbar \omega_{f}}{2}\right),
\end{equation}
\begin{equation}
    \label{eq:H-11}d_{i}=\frac{\omega_{i}}{\hbar}\tanh \left(\frac{\tau_{i}\hbar \omega_{i}}{2}\right), \quad d_{f}=\frac{\omega_{f}}{\hbar}\tanh \left(\frac{\tau_{f}\hbar \omega_{f}}{2}\right),
\end{equation}

Based on these definitions, two matrices are further defined,
\begin{equation}
    \label{eq:H-12}C=c_{i}+c_{f}, \quad D=d_{i}+d_{f}
\end{equation}

Under the Condon approximation, the Fourier-transformed (time-dependent) lineshape function can be expressed as,
\begin{equation}
    \label{eq:H-13}\chi(t)=\sqrt{\frac{a_{i}a_{f}}{(i \hbar)^{2}(C D)}}\exp \left(\Delta Q^{2}d_{i}+D^{-1}\Delta Q^{2}d_{i}^{2}\right)
\end{equation}

The non-Condon effect can be viewed as a correction to Eq.~\eqref{eq:H-13}. We define the non-Condon term as,
\begin{equation}
    \label{eq:H-14}A(t)=\frac{1}{2}\left(D^{-1}-C^{-1}\right)+\left(D^{-1}\Delta Q d_{i}\right)^{2}
\end{equation}

Finally, including the non-Condon effect,
\begin{equation}
    \label{eq:H-15}\chi_{HT}(t)=\chi(t) A(t).
\end{equation}

A single integration of Eq.~\eqref{eq:H-15} from $-\infty$ to $+\infty$ yields the result of lineshape function and the transition rate without the need for summation over $m$ and $n$.

\begin{figure}[htbp!]
    \centering
    \includegraphics[width=0.49\textwidth]{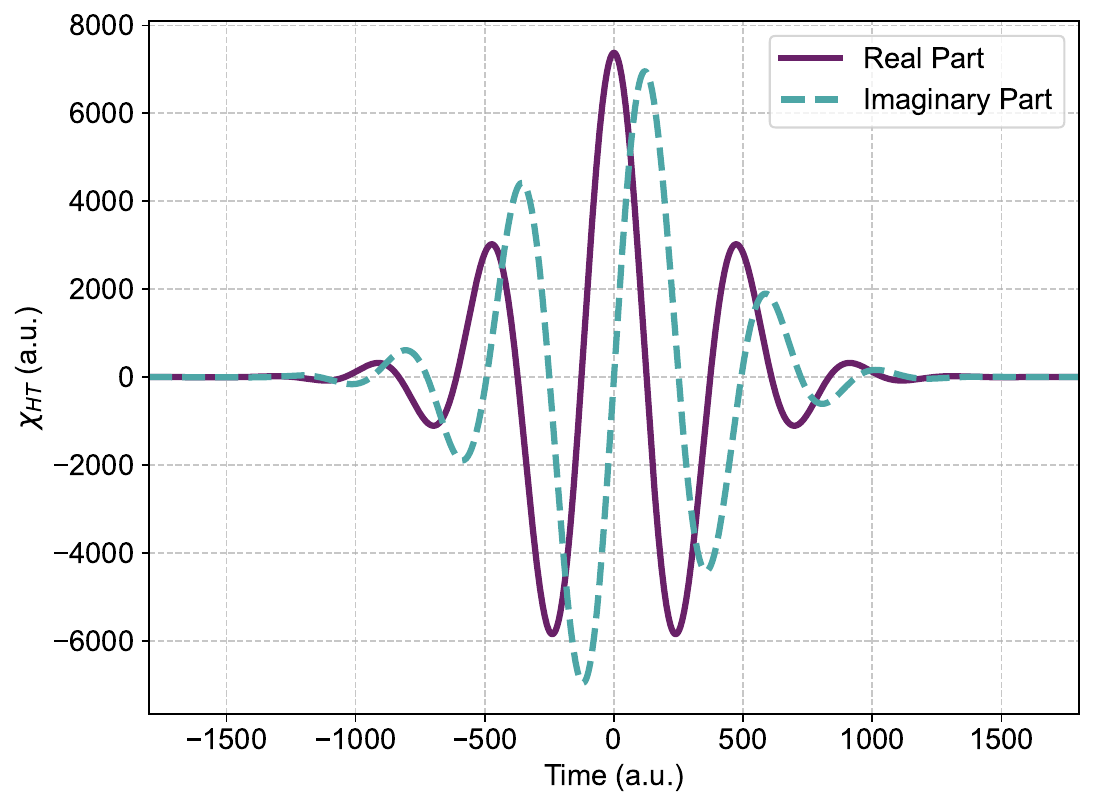}
    \caption{Time-dependent $\chi_{HT}$(t) including non-Condon effects in the Fourier transform method.}
    \label{Fig:H-2}
\end{figure}

Substituting the configuration coordinate parameters into Eq.~\eqref{eq:H-8}-\eqref{eq:H-15}, we can plot $\chi_{HT}(t)$ as a function of time $t$, as shown in Fig.~\ref{Fig:H-2}. Both the real and imaginary parts approach zero when $|t|$ exceeds 200 a.u. Compared to energy-space methods, the Fourier transform approach significantly accelerates calculations and requires no specification of $m$ and $n$, making it well-suited for repeated transition rate calculations in device simulations.

It is worth noting that we can further accelerate the calculation based on the above Fourier transform method by considering the specific operating conditions of the device. Specifically, when a MOSFET is in operation, the gate voltage may vary with time. As show in Eq.~\eqref{eq:4}, the gate voltage affects carrier capture at defect in two aspects. First, changes in gate voltage affect the voltage drop across the oxide layer and surface potential, thereby causing the defect energy level to shift relative to the semiconductor band edge. Second, changes in gate voltage affect the surface potential of the semiconductor, which in turn affects the carrier concentration in the channel. According to the capture rate formula $k_{c}=C_{p}p$, the former affects $C_{p}$ while the latter affects $p$, so both factors influence the carrier capture process. We first discuss the effect of defect level variation on the capture coefficient $C_{p}$ (transition rate $r$).

Since the oxide is typically amorphous, different defect configurations exhibit structural variations, resulting in slightly different $\Delta Q$ values (Eq.~\eqref{eq:H-6}). When the gate voltage changes: (i) the defect level shift causes a change in transition energy $\Delta E$, resulting in vertical displacement of the two potential energy surfaces in the configuration coordinate diagram; (ii) different $\Delta Q$ values cause horizontal displacement of the potential energy surfaces. Therefore, by traversing all combinations of ($\Delta E$, $\Delta Q$) and computing their lineshape functions using the Fourier transform, transition rates can be calculated. We propose an interpolation method for the lineshape function as follows: (i) firstly, we establish a coarse grid in the ($\Delta E$, $\Delta Q$) two-dimensional space and rapidly compute the lineshape functions at these grid points using the Fourier transform method in parallel; (ii) secondly, we use two-dimensional spline interpolation to obtain lineshape function values on a finer grid, generating a continuous surface over the ($\Delta E$, $\Delta Q$) space; (iii) thirdly, for a specific defect type (with definite $\Delta E$ and $\Delta Q$ values), we substitute ($\Delta E$, $\Delta Q$) into the two-dimensional surface to quickly obtain the lineshape function value.

\begin{figure}[htbp!]
    \centering
    \includegraphics[width=0.49\textwidth]{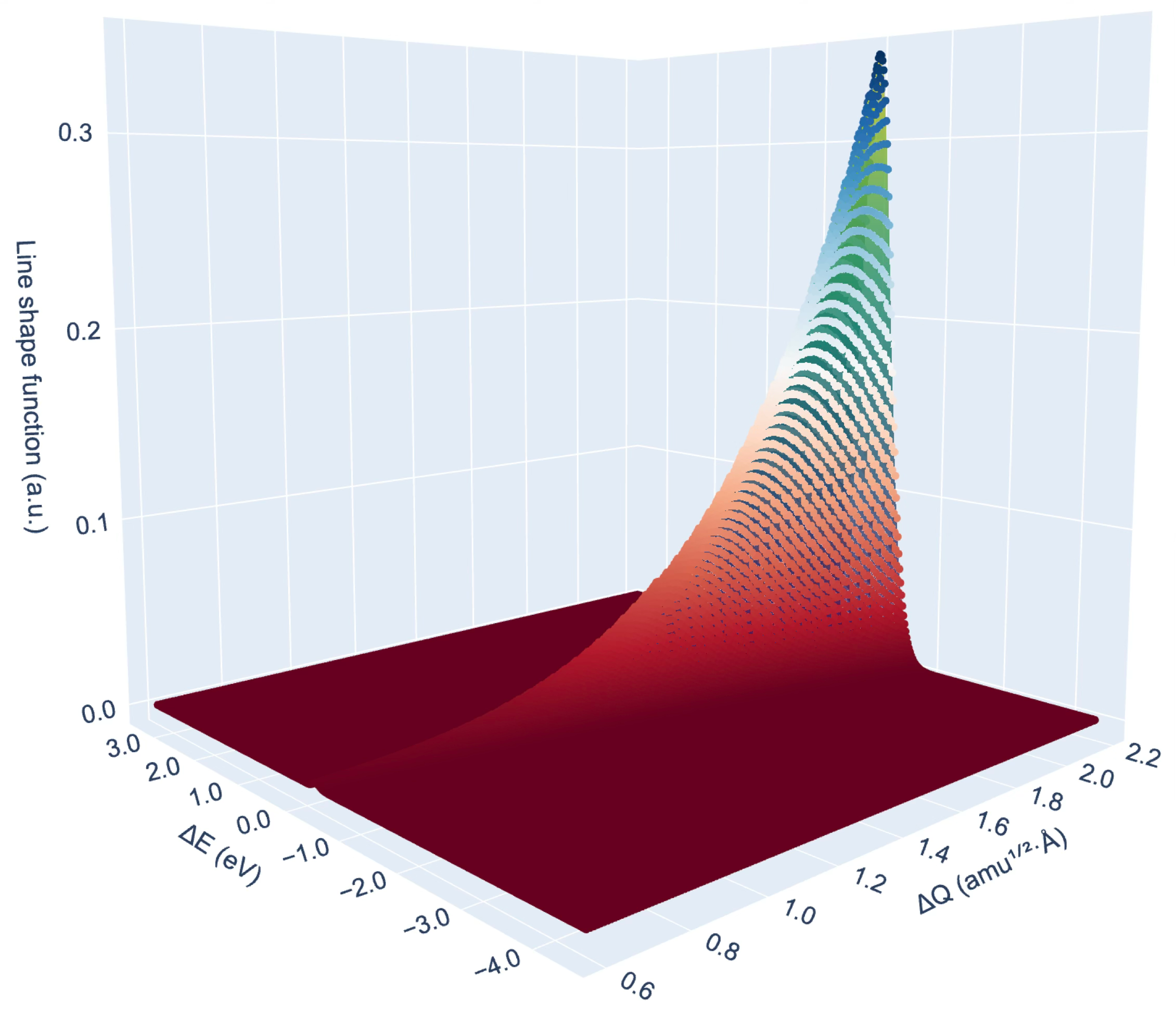}
    \caption{\red{Accuracy test of the lineshape-function interpolation method. The surface shows the lineshape function obtained by interpolation as a function of $(\Delta E,\Delta Q)$, while the markers denote values computed directly. The markers lie on the interpolated surface, demonstrating the high accuracy of the interpolation method.}}
    \label{Fig:H-3}
\end{figure}

Validation shows that this interpolation method efficiently computes transition rates for tens of thousands of defects (10,000 defects in 87 ms) with high accuracy, as shown in Fig.~\ref{Fig:H-3}: the surface represents the continuous lineshape function, while the points indicate values calculated precisely using the Fourier transform at these grid points. These points fall exactly on the surface, demonstrating the accuracy of this method. Therefore, in subsequent simulations, we use the Fourier transform combined with two-dimensional interpolation for transition rate calculations.

It is worth noting that, when combined with the lineshape function interpolation method, we can directly solve the Schrödinger equation to calculate the lineshape function instead of using the Fourier transform method. The main advantage of this approach is that it allows us to consider anharmonic potential energy surfaces. By combining direct solution of the Schrödinger equation with two-dimensional interpolation, anharmonic effects can be incorporated into the lineshape function, which is particularly important for systems with large lattice relaxation.
	
	\subsubsection{Electronic overlap integral}
	Alkauskas et al. provided a detailed method for calculating $W_{if}=\left\langle\psi_{i}\left| \frac{\partial H_{eL}}{\partial Q}\right|\psi_{f}\right\rangle$~\cite{RN26}. Under the single-mode approximation, the electronic wavefunction overlap integral can be computed using density functional theory (DFT). It should be noted that this method requires the wavefunctions $\psi_{i}$ and $\psi_{f}$ to be defined within the same periodic supercell, i.e., the initial-state wavefunction (typically the semiconductor band edge state) and the final-state wavefunction (the defect state) must be located at the same spatial position. However, this is not applicable to electronic wavefunction overlap calculations in electronic devices, since the final state wavefunction (defect state) is located in the oxide while the initial state wavefunction (semiconductor band edge) is at the semiconductor channel, with a spatial separation between them. Consequently, calculating the overlap integral using DFT single-particle wavefunctions presents certain difficulties.

    To relate the electronic wavefunction calculation to the defect parameters $\Delta E$ and $\Delta Q$ in the lineshape function, as well as to the localization of the defect state, \redcnew{following Ref.~\cite{RN26}} we approximate the electronic wavefunction overlap integral as,
\begin{equation}
    \label{eq:H-16}W_{if}\approx \frac{\Delta E}{\Delta Q}\sqrt{\frac{N_{\text{local}}}{N_{\text{all}}}},
\end{equation}
{where $N_{\mathrm{local}}$ is the number of atoms covered by the spatial distribution of the final state wavefunction $\psi_{f}$, and $N_{\mathrm{all}}$ is the total number of atoms in the (semiconductor) system.} Note that the $W_{if}$ value calculated using Eq.~\eqref{eq:H-16} is typically an upper bound, since this equation does not account for the atomic orbital composition and spatial symmetry of $\psi_{i}$ and $\psi_{f}$; including these factors would further reduce the value. The term $N_{\mathrm{all}}$ in Eq.~\eqref{eq:H-16} lacks a quantitative standard, i.e., $N_{\mathrm{all}}$ increases with the system size, making the overlap integral value entirely dependent on system size. 

We use the system volume to cancel this size dependence, so the calculated electronic wavefunction overlap integral no longer depends on system size but is related to the semiconductor lattice density.

Furthermore, since $\psi_{i}$ and $\psi_{f}$ are spatially separated, their overlap integral decreases rapidly with increasing distance. We use the WKB tunneling factor (as shown in Eq.~\eqref{WKB}) to describe this effect. Combining all of the above, the transition rate from Eq.~\eqref{eq:H-4} can be directly written as the capture coefficient when volume is considered,
\begin{equation}\label{eq:H-18}
\begin{aligned}
    C_{p}&=\frac{2 \pi}{\hbar}f_\text{WKB}V\left|W_{if}\right|^{2}G\\
    &=Vrf_\text{WKB}.
\end{aligned}
\end{equation}

The second effect of gate voltage variation on the carrier capture process is the change in carrier concentration in the channel. Following the analysis in Sec. \ref{sec:detailedbalance}, the capture rate can be written as $k_{c}=C_{p}p$, and the emission rate as $k_{e}=C_{p}N_{v}\exp{(-\frac{E_{T}-E_{v}}{k_{B}T})}$. Therefore, this effect is already incorporated.
	\redc{\subsection{Thermal transition rate}}\label{sec:Thermal}
	Thermal transitions between different defect configurations in same charge state are described within classical transition-state theory (TST)~\cite{TST1, TST2, TST3, TST4},
	\begin{equation}\label{eq:thermal}
    k_{\mathrm{T}} = \nu\,\exp\left(-\frac{E_b}{k_{\mathrm{B}}T}\right),
\end{equation}
where $\nu$ is the attempt frequency and $E_b$ is the transition barrier.
	
    \ 
	
    \section{ALL-STATE MODEL}\label{sec:all-state-model}
	
	In real electronic devices, gate dielectrics are typically amorphous oxides, such as a-SiO$_{\mathrm{2}}$, a-HfO$_{\mathrm{2}}$, and a-Al$_{\mathrm{2}}$O$_{\mathrm{3}}$ \cite{Huang10, XiangweiJiang2025}. Due to their low symmetry and complex potential energy surfaces, defects in these materials exhibit highly complex structures \cite{Shluger2020, Strand_2018, XiangweiJiang2025, guo2024all, WILHELMER2022114801}. Taking the oxygen vacancy (V$_{\mathrm{O}}$) in a-SiO$_{\mathrm{2}}$ as an example, in previous work \cite{guo2024all}, we performed high-throughput, near-global structural searches at different oxygen sites using Defect and Dopant ab-initio Simulation Package (DASP \cite{Huang_2022}). We applied various random structural perturbations around V$_{\mathrm{O}}$ and employed a near-global search method. The results show that neutral V$_{\mathrm{O}}$ in a-SiO$_{\mathrm{2}}$ mainly adopts four configurations: Si-dimer, left-back-projected, right-back-projected, and double-back-projected. The positively charged V$_{\mathrm{O}}^{+}$ not only adopts these four configurations but also forms three additional ones: left-in-plane, right-in-plane, and twisted, giving a total of seven possible configurations. Depending on the unique local environment of each oxygen site, any of these configurations can be the ground state of V$_{\mathrm{O}}^{+}$.

The complexity of defect structures and energetics in amorphous gate dielectrics poses significant challenges for accurately describing their impact on device reliability. To address this complexity, we proposed the ``all-state model'', which systematically considers the all structural and charge states of defects in amorphous oxides, as well as all possible transitions among these states. These transitions can be classified into two categories based on whether the charge state changes. (i) When a defect captures or emits a carrier, its charge state changes. This type of transition is a nonradiative multiphonon (NMP) transition. (ii) When no carrier capture or emission occurs, the charge state remains unchanged and only the structural configuration changes. This type of transition is a thermal transition.

\begin{figure}[t]
        \centering
        \includegraphics[width=0.48\textwidth]{
            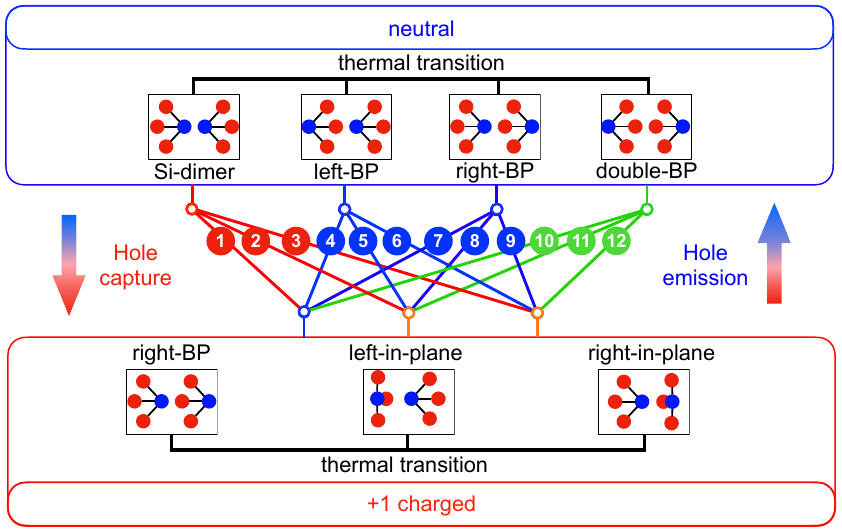
        }
        \caption{\red{Stable and metastable configurations of V$_\mathrm{O}^{0}$ and V$_\mathrm{O}^{+}$ formed at the selected oxygen site, and the corresponding transition pathways. 
In the neutral charge state, V$_\mathrm{O}^{0}$ exists four configurations, i.e., Si-dimer, left-back-projected (left-BP), right-back-projected (right-BP), and double-back-projected (double-BP), where the Si-dimer configuration is the ground-state. In the $+1$ charge state, V$_\mathrm{O}^{+}$ exhibits three configurations, i.e., left-in-plane, right-in-plane, and right-back-projected (right-BP), where the left-in-plane configuration is the ground-state. 
Thermal transitions occur between configurations with the same charge state and are denoted by the horizontal black lines. NMP transitions (hole capture/emission) occur between configurations of different charge states and are denoted by the red, blue, and green lines.}}
        \label{Fig:QQP_Structure}
    \end{figure}

Next, we use V$_{\mathrm{O}}$ at a single oxygen site in a-SiO$_{\mathrm{2}}$ as an example to illustrate how the all-state model accounts for multiple transition pathways during hole capture and emission, and their effects on device threshold voltage. As shown in Fig.~\ref{Fig:QQP_Structure}, for V$_{\mathrm{O}}$ formed at this oxygen site, neutral V$_{\mathrm{O}}$ has four stable configurations: Si-dimer, left-back-projected, right-back-projected, and double-back-projected. Among these, the Si-dimer configuration is the ground state. When neutral V$_{\mathrm{O}}$ captures a hole and becomes V$_{\mathrm{O}}^{+}$, three stable configurations exist at this site: left-in-plane, right-in-plane, and right-back-projected. Among these, the right-in-plane configuration is the ground state. During hole capture, neutral V$_{\mathrm{O}}$ in any configuration can transform into any of the three V$_{\mathrm{O}}^{+}$ configurations via NMP transitions. For example, V$_{\mathrm{O}}$ in the Si-dimer configuration can transform into V$_{\mathrm{O}}^{+}$ in the left-in-plane, right-in-plane, or right-back-projected configuration. Similarly, V$_{\mathrm{O}}$ in the left-back-projected, right-back-projected, and double-back-projected configurations can also undergo these three transitions, resulting in a total of 12 possible NMP transitions for hole capture. Likewise, 12 possible NMP transitions exist for hole emission. When V$_{\mathrm{O}}$ transforms between different configurations within the same charge state without carrier exchange, the process is a thermal transition. For example, V$_{\mathrm{O}}$ in the Si-dimer configuration can transform into the left-back-projected, right-back-projected, or double-back-projected configuration. The other three neutral configurations can similarly transform into one another, giving 6 possible thermal transitions in total. Similarly, the three V$_{\mathrm{O}}^{+}$ configurations can transform among themselves, yielding 3 possible thermal transitions. All these NMP and thermal transitions compete with each other, collectively influencing the carrier capture and emission dynamics of defects and their charge states. Therefore, the all-state model considers all possible NMP and thermal transitions when simulating the impact of defect-related carrier capture and emission on device reliability. Furthermore, the possible configurations and relative stabilities of V$_{\mathrm{O}}$ and V$_{\mathrm{O}}^{+}$ vary among different oxygen sites \cite{guo2024all}. Accurate simulation thus requires incorporating all these site-dependent differences.

Considering all oxygen sites in a-SiO$_{\mathrm{2}}$, neutral V$_{\mathrm{O}}$ exhibits four possible configurations (Si-dimer, left-back-projected, right-back-projected, and double-back-projected), while V$_{\mathrm{O}}^{+}$ exhibits seven possible configurations across different oxygen sites (including the above four as well as left-in-plane, right-in-plane, and twisted). By considering all possible transition pathways at each oxygen site, the all-state model enables accurate description of the complex dynamics of defects in amorphous oxides and their impact on device reliability.

	\section{MASTER EQUATIONS OF ALL-STATE MODEL}\label{sec:master_eq}
	
	As described in Sec.~\ref{sec:all-state-model}, due to the inherent low symmetry of amorphous oxides, certain defects within them can adopt multiple configurations in a given charge state. Since these configurations have similar formation energies, all of them must be considered when simulating the impact of defects on device reliability. This section describes how to solve for the time evolution of the probabilities of defect in each state based on the all-state model.
Consider a donor defect $\alpha$ in an amorphous oxide. Before capturing a hole, the defect is in its neutral state ($\alpha^{0}$); after capturing a hole, it transforms into the +1 charged state ($\alpha^{+}$). Assume that $\alpha^{0}$ has 2 configurations and $\alpha^{+}$ also has 2 configurations (for a detailed discussion of the generalization to $M$ configurations for $\alpha^{0}$ and $N$ configurations for $\alpha^{+}$, see Appendix; this section focuses on the evolution equations and their solution methods). Assume the probabilities of defect $\alpha$ in each state at time $t=t_{0}$ are,

\begin{widetext}
\begin{equation}
    \label{eq:initial_condition}\mathbf{P}(t_{0}) = \left(P_{\mathrm{neutral}}^{\mathrm{struc}-1}(t_{0}),\ P_{\mathrm{neutral}}^{\mathrm{struc}-2}(t_{0}),\ P_{\mathrm{charged}}^{\mathrm{struc}-1}(t_{0}),\ P_{\mathrm{charged}}^{\mathrm{struc}-2}(t_{0})\right)
\end{equation}
\end{widetext}

The time evolution of the probabilities of defect $\alpha$ in each state during the time interval $[t_{0}, t_{0}+ \Delta t]$ satisfies the following equations,

\begin{widetext}
{
\begin{equation}
    \begin{aligned}
        \label{eq:all_state_master_equation_1} \small
        & \frac{d P_{\mathrm{neutral }}^{\mathrm{struc}-1}}{d t}= -P_{\mathrm{neutral }}^{\mathrm{struc}-1}\cdot\left(k_{c}^{\mathrm{n} 1 \rightarrow \mathrm{c} 1}+k_{c}^{\mathrm{n} 1 \rightarrow \mathrm{c} 2}+k_\text{T}^{\mathrm{n} 1 \rightarrow \mathrm{n} 2}\right)+P_{\mathrm{neutral }}^{\mathrm{struc}-2}\cdot k_\text{T}^{\mathrm{n} 2 \rightarrow \mathrm{n} 1}+P_{\mathrm{charged}}^{\mathrm{struc}-1}\cdot k_{e}^{\mathrm{c} 1 \rightarrow \mathrm{n} 1}+P_{\mathrm{charged }}^{\mathrm{struc}-2}\cdot k_{e}^{\mathrm{c} 2 \rightarrow \mathrm{n} 1} \\
        & \frac{d P_{\mathrm{neutral }}^{\mathrm{struc}-2}}{d t}= P_{\mathrm{neutral }}^{\mathrm{struc}-1}\cdot k_\text{T}^{\mathrm{n} 1 \rightarrow \mathrm{n} 2}-P_{\mathrm{neutral }}^{\mathrm{struc}-2}\cdot\left(k_{c}^{\mathrm{n} 2 \rightarrow \mathrm{c} 1}+k_{c}^{\mathrm{n} 2 \rightarrow \mathrm{c} 2}+k_\text{T}^{\mathrm{n} 2 \rightarrow \mathrm{n} 1}\right)+P_{\mathrm{charged}}^{\mathrm{struc}-1}\cdot k_{e}^{\mathrm{c} 1 \rightarrow \mathrm{n} 2}+P_{\mathrm{charged}}^{\mathrm{struc}-2}\cdot k_{e}^{\mathrm{c} 2 \rightarrow \mathrm{n} 2}   \\
        & \frac{d P_{\mathrm{charged }}^{\mathrm{struc}-1}}{d t}= P_{\mathrm{neutral }}^{\mathrm{struc}-1}\cdot k_{c}^{\mathrm{n} 1 \rightarrow \mathrm{c} 1}+P_{\mathrm{neutral }}^{\mathrm{struc}-2}\cdot k_{c}^{\mathrm{n} 2 \rightarrow \mathrm{c} 1}-P_{\mathrm{charged}}^{\mathrm{struc} 1}\cdot\left(k_{e}^{\mathrm{c} 1 \rightarrow \mathrm{n} 2}+k_{e}^{\mathrm{c} 1 \rightarrow \mathrm{n} 1}+k_\text{T}^{\mathrm{c} 1 \rightarrow \mathrm{c} 2}\right)+P_{\mathrm{charged}}^{\mathrm{struc}-2}\cdot k_\text{T}^{\mathrm{c} 2 \rightarrow \mathrm{c} 1}   \\
        & \frac{d P_{\mathrm{charged}}^{\mathrm{struc}-2}}{d t}= P_{\mathrm{neutral}}^{\mathrm{struc}-1}\cdot k_{c}^{\mathrm{n} 1 \rightarrow \mathrm{c} 2}+P_{\mathrm{neutral }}^{\mathrm{struc}-2}\cdot k_{c}^{\mathrm{n} 2 \rightarrow \mathrm{c} 2}+P_{\mathrm{charged}}^{\mathrm{struc}-1}\cdot k_\text{T}^{\mathrm{c} 1 \rightarrow \mathrm{c} 2}-P_{\mathrm{charged}}^{\mathrm{struc}-2}\cdot\left(k_{e}^{\mathrm{c} 2 \rightarrow \mathrm{n} 1}+k_{e}^{\mathrm{c} 2 \rightarrow \mathrm{n} 2}+k_\text{T}^{\mathrm{c} 2 \rightarrow \mathrm{c} 1}\right)
    \end{aligned}
\end{equation}
}
\end{widetext}

Here, $P_{\mathrm{neutral}}^{\mathrm{struc}-i}$ denotes the probability of the defect in neutral configuration $i$, and $P_{\mathrm{charged}}^{\mathrm{struc}-i}$ denotes the probability of the defect in charged configuration $i$. The rate $k_{c}^{n1\rightarrow c1}$ represents the capture rate at which a defect in neutral configuration 1 ($n1$) captures a carrier and transforms into charged configuration 1 ($c1$). The rate $k_{e}^{c1\rightarrow n1}$ represents the emission rate at which a defect in charged configuration 1 ($c1$) emits a carrier and transforms into neutral configuration 1 ($n1$).
The rate $k_\text{T}^{\mathrm{n1\rightarrow n2}}$ represents the thermal transition rate at which a defect in neutral configuration $\mathrm{n1}$ transforms into configuration $\mathrm{n2}$ within the same charge state.

For convenience in formulation and solution, 
we define the continuous-time Markov chain generator matrix as,

\begin{widetext}
\begin{equation}
\label{eq:A-matrix}
\begin{aligned}
&\mathbf{A}(V_\mathrm{G}, T) = \\
&\begin{pmatrix}
-\left(\sum_{i=1}^{2} k_{c}^{\mathrm{n}1 \to \mathrm{c}i} + k_\text{T}^{\mathrm{n}1 \to \mathrm{n}2}\right) & 
k_\text{T}^{\mathrm{n}2 \to \mathrm{n}1} & 
k_{e}^{\mathrm{c}1 \to \mathrm{n}1} & 
k_{e}^{\mathrm{c}2 \to \mathrm{n}1} \\
k_\text{T}^{\mathrm{n}1 \to \mathrm{n}2} & 
-\left(\sum_{i=1}^{2} k_{c}^{\mathrm{n}2 \to \mathrm{c}i} + k_\text{T}^{\mathrm{n}2 \to \mathrm{n}1}\right) & 
k_{e}^{\mathrm{c}1 \to \mathrm{n}2} & 
k_{e}^{\mathrm{c}2 \to \mathrm{n}2} \\
k_{c}^{\mathrm{n}1 \to \mathrm{c}1} & 
k_{c}^{\mathrm{n}2 \to \mathrm{c}1} & 
-\left(\sum_{j=1}^{2} k_{e}^{\mathrm{c}1 \to \mathrm{n}j} + k_\text{T}^{\mathrm{c}1 \to \mathrm{c}2}\right) & 
k_\text{T}^{\mathrm{c}2 \to \mathrm{c}1} \\
k_{c}^{\mathrm{n}1 \to \mathrm{c}2} & 
k_{c}^{\mathrm{n}2 \to \mathrm{c}2} & 
k_\text{T}^{\mathrm{c}1 \to \mathrm{c}2} & 
-\left(\sum_{j=1}^{2} k_{e}^{\mathrm{c}2 \to \mathrm{n}j} + k_\text{T}^{\mathrm{c}2 \to \mathrm{c}1}\right)
\end{pmatrix}^{\!T}.
\end{aligned}
\end{equation}
\end{widetext}

The evolution Eq.~\eqref{eq:all_state_master_equation_1} for the probabilities of defect $\alpha$ in each state during the time interval $[t_{0}, t_{0}+ \Delta t]$ can then be expressed as:
\begin{equation}
    \label{eq:ODES-2}\frac{d}{dt}\mathbf{P}(t) = \mathbf{P}(t) \cdot \mathbf{A}(V_\mathrm{G}, T)\ .
\end{equation}

In particular, when $t_{0}=0$, no bias voltage has been applied to the device, and the device is in steady state. At this point, $\frac{d\mathbf{P}}{dt} = 0$, which gives:
\begin{equation}
    \label{t=0_condition}\mathbf{P}(t=0)\mathbf{A}(V_\mathrm{G}^\prime, T^{\prime})=\mathbf{0}
\end{equation}
where $V_\mathrm{G}^\prime$ and $T^{\prime}$ are the gate voltage and temperature at $t=0$. \redcnew{Since $\mathbf{A}(V_\mathrm{G}^\prime,T^\prime)$ is a CTMC generator matrix with all off-diagonal entries $k>0$, the associated directed graph is strongly connected, which implies that the generator matrix $\mathbf{A}(V_\mathrm{G}^\prime,T^\prime)$ is irreducible~\cite{Norris1997}. For a finite-state CTMC, irreducibility implies that the state space forms a single closed communicating class~\cite{Norris1997}, so the stationary subspace is one-dimensional, i.e., $\dim\ker(\mathbf{A}^T)=1$~\cite{Norris1997}. By the rank--nullity theorem, $\mathrm{rank}(\mathbf{A})=4-1$~\cite{HornJohnson2012}. Thus, Eq.~\eqref{t=0_condition} exists a unique solution under the normalization $\mathbf{P}(t=0)\mathbf{1}^T=1$.} To obtain the probabilities of the defect in each state when the device is in steady state, we need to solve the linear system Eq.~\eqref{t=0_condition}. From Eq.~\eqref{eq:A-matrix}, we observe that $|A| = 0$. Meanwhile, the total probability of the defect in all configurations is conserved and equals 1 throughout the carrier capture and emission process:
\begin{widetext}
\begin{equation}
    P_{\mathrm{neutral }}^{\mathrm{struc}-1}(t=0)+P_{\mathrm{neutral }}^{\mathrm{struc}-2}(t=0)+P_{\mathrm{charged }}^{\mathrm{struc}-1}(t=0)+P_{\mathrm{charged }}^{\mathrm{struc}-2}(t=0)=1
\end{equation}
\end{widetext}

Therefore, we perform Gaussian elimination on the linear system Eq.~\eqref{t=0_condition}, expressing $P_{\mathrm{neutral }}^{\mathrm{struc}-1}(t=0)$ in terms of the other three probabilities. Substituting this into Eq.~\eqref{t=0_condition}, we obtain a system of equations for $P_{\mathrm{neutral }}^{\mathrm{struc}-2}(t=0)$, $P_{\mathrm{charged }}^{\mathrm{struc}-1}(t=0)$, and $P_{\mathrm{charged }}^{\mathrm{struc}-2}(t=0)$:

\begin{widetext}
\centering
\small 
\begin{equation*}
    \begin{aligned}
        0 &= -{P'}_{\mathrm{neutral}}^{\mathrm{struc}-2} \left( \sum_{i=1}^{2} {k'}_{c}^{\mathrm{n}2 \to \mathrm{c}i} + {k'}_{T}^{\mathrm{n} 2 \to \mathrm{n} 1} + {k'}_{T}^{\mathrm{n}1 \to \mathrm{n}2} \right) 
             +{P'}_{\mathrm{charged}}^{\mathrm{struc}-1} \left( {k'}_{e}^{\mathrm{c} 1 \to \mathrm{n} 2} - {k'}_{T}^{\mathrm{n} 1 \to \mathrm{n} 2} \right) 
             +{P'}_{\mathrm{charged}}^{\mathrm{struc}-2} \left( {k'}_{e}^{\mathrm{c} 2 \to \mathrm{n} 2} - {k'}_{T}^{\mathrm{n} 1 \to \mathrm{n} 2} \right) 
             +{k'}_{T}^{\mathrm{n} 1 \to \mathrm{n} 2} \\
        0 &= {P'}_{\mathrm{neutral}}^{\mathrm{struc}-2} \left( {k'}_{c}^{\mathrm{n} 2 \to \mathrm{c} 1} - {k'}_{c}^{\mathrm{n} 1 \to \mathrm{c} 1} \right) 
             -{P'}_{\mathrm{charged}}^{\mathrm{struc}-1} \left( \sum_{j=1}^{2} {k'}_{e}^{\mathrm{c} 1 \to \mathrm{n} j} + {k'}_{T}^{\mathrm{c} 1 \to \mathrm{c} 2} + {k'}_{c}^{\mathrm{n} 1 \to \mathrm{c} 1} \right) 
             +{P'}_{\mathrm{charged}}^{\mathrm{struc}-2} \left( {k'}_{T}^{\mathrm{c} 2 \to \mathrm{c} 1} - {k'}_{c}^{\mathrm{n} 1 \to \mathrm{c} 1} \right) 
             +{k'}_{c}^{\mathrm{n} 1 \to \mathrm{c} 1} \\
        0 &= {P'}_{\mathrm{neutral}}^{\mathrm{struc}-2} \left( {k'}_{c}^{\mathrm{n} 2 \to \mathrm{c} 2} - {k'}_{c}^{\mathrm{n} 1 \to \mathrm{c} 2} \right) 
             +{P'}_{\mathrm{charged}}^{\mathrm{struc}-1} \left( {k'}_{T}^{\mathrm{c} 1 \to \mathrm{c} 2} - {k'}_{c}^{\mathrm{n} 1 \to \mathrm{c} 2} \right) 
             -{P'}_{\mathrm{charged}}^{\mathrm{struc}-2} \left( \sum_{j=1}^{2} {k'}_{e}^{\mathrm{c} 2 \to \mathrm{n} j} + {k'}_{T}^{\mathrm{c} 2 \to \mathrm{c} 1} + {k'}_{c}^{\mathrm{n} 1 \to \mathrm{c} 2} \right) 
             +{k'}_{c}^{\mathrm{n} 1 \to \mathrm{c} 2}
    \end{aligned}
\end{equation*}
\end{widetext}

Here, ${P^{\prime}}_{\mathrm{neutral}}^{\mathrm{struc}-i}$ and ${P^{\prime}}_{\mathrm{charged}}^{\mathrm{struc}-i}$ denote the occupation probabilities of neutral and charged defect configurations at $t=0$, respectively, and ${k^{\prime}}_{c/e/T}^{i \rightarrow k}$ denotes the transition rate from state $i$ to state $k$ at $t=0$. The above system can be written in matrix form:
\begin{equation}
    \mathbf{0}=\mathbf{P^\prime}(t=0) \cdot \mathbf{A^\prime}(V_\mathrm{G}^\prime, T^{\prime}) + \mathbf{f}
\end{equation}

where:

\begin{equation*}
    \mathbf{P^\prime}(t=0) = \left({P^{\prime}}_{\mathrm{neutral }}^{\mathrm{struc}-2},{P^{\prime}}_{\mathrm{charged }}^{\mathrm{struc}-1},{P^{\prime}}_{\mathrm{charged}}^{\mathrm{struc}-2}\right)
\end{equation*}

\begin{equation*}
    \mathbf{f}= \left({k^{\prime}}_{T}^{\mathrm{n} 1 \rightarrow \mathrm{n} 2},{k^{\prime}}_{c}^{\mathrm{n} 1 \rightarrow \mathrm{c} 1},{k^{\prime}}_{c}^{\mathrm{n} 1 \rightarrow \mathrm{c} 2}\right)
\end{equation*}

\begin{widetext}
\centering
\begin{equation*}
\begin{aligned}
    &\mathbf{A'}(V_\mathrm{G}', T') =\\
    &\begin{pmatrix}
        -\sum_{i=1}^{2} {k'}_{c}^{\mathrm{n}2 \to \mathrm{c}i} - {k'}_{T}^{\mathrm{n} 2 \to \mathrm{n} 1} - {k'}_{T}^{\mathrm{n} 1 \to \mathrm{n} 2} & 
        {k'}_{e}^{\mathrm{c} 1 \to \mathrm{n} 2} - {k'}_{T}^{\mathrm{n} 1 \to \mathrm{n} 2} & 
        {k'}_{e}^{\mathrm{c} 2 \to \mathrm{n} 2} - {k'}_{T}^{\mathrm{n} 1 \to \mathrm{n} 2} \\
        {k'}_{c}^{\mathrm{n} 2 \to \mathrm{c} 1} - {k'}_{c}^{\mathrm{n} 1 \to \mathrm{c} 1} & 
        -\sum_{j=1}^{2} {k'}_{e}^{\mathrm{c} 1 \to \mathrm{n} j} - {k'}_{T}^{\mathrm{c} 1 \to \mathrm{c} 2} - {k'}_{c}^{\mathrm{n} 1 \to \mathrm{c} 1} & 
        {k'}_{T}^{\mathrm{c} 2 \to \mathrm{c} 1} - {k'}_{c}^{\mathrm{n} 1 \to \mathrm{c} 1} \\
        {k'}_{c}^{\mathrm{n} 2 \to \mathrm{c} 2} - {k'}_{c}^{\mathrm{n} 1 \to \mathrm{c} 2} & 
        {k'}_{T}^{\mathrm{c} 1 \to \mathrm{c} 2} - {k'}_{c}^{\mathrm{n} 1 \to \mathrm{c} 2} & 
        -\sum_{j=1}^{2} {k'}_{e}^{\mathrm{c} 2 \to \mathrm{n} j} - {k'}_{T}^{\mathrm{c} 2 \to \mathrm{c} 1} - {k'}_{c}^{\mathrm{n} 1 \to \mathrm{c} 2}
    \end{pmatrix}^{\!T}
\end{aligned}
\end{equation*}
\end{widetext}

\redcnew{Therefore, after eliminating $P_{\mathrm{neutral}}^{\mathrm{struc}-1}(t=0)$ using the probability conservation constraint, Eq.~\eqref{t=0_condition} reduces to the $3\times3$ linear system $\mathbf{0}=\mathbf{P^\prime}(t=0)\cdot\mathbf{A^\prime}(V_\mathrm{G}^\prime,T^\prime)+\mathbf{f}$. Since the original CTMC $\mathbf{A}(V_\mathrm{G}^\prime,T^\prime)$ is irreducible, the stationary subspace of $\mathbf{A}^T(V_\mathrm{G}^\prime,T^\prime)$ is one-dimensional, and imposing the normalization constraint removes the null degree of freedom. Equivalently, the three equations obtained after Gaussian elimination are linearly independent, so $\mathbf{A^\prime}(V_\mathrm{G}^\prime,T^\prime)$ is a $3\times3$ matrix with $\mathrm{rank}(\mathbf{A^\prime})=3$, and is therefore nonsingular, i.e., invertible. The unique solution is given by,}

\begin{equation}
    \label{eq:steady}\mathbf{P^\prime}(t=0) = - \mathbf{f}\cdot\mathbf{A^\prime}(V_\mathrm{G}^\prime, T^{\prime})^{-1}
\end{equation}

Eq.~\eqref{eq:steady} gives the probabilities of defect $\alpha$ in each state when the device is in steady state at $t=0$. For $t>0$, depending on the voltage applied to the gate, defect $\alpha$ will exchange carriers with the channel and the gate.  
According to the analysis in Sec.~\ref{sec:capture_and_emission_rate}, NMP transition rates depend on $V_{\mathrm{G}}$ and temperature $T$. When the gate voltage $V_{\mathrm{G}}$ and temperature $T$ remain constant, all matrix elements in Eq.~\eqref{eq:A-matrix} are constants, and thus $\mathbf{A}(V_\mathrm{G}, T)$ is a constant matrix. According to Appendix, the solution to the differential Eq.~\eqref{eq:ODES-2} with initial condition Eq.~\eqref{eq:initial_condition} exists and is unique. When the gate voltage $V_{\mathrm{G}}$ and temperature $T$ vary with time, we take $\Delta t$ to be infinitesimally small, so that $V_{\mathrm{G}}$ can be approximated as constant during the time interval $[t_{0}, t_{0} + \Delta t]$. In this case, $\mathbf{A}(V_\mathrm{G}, T)$ can still be treated as a constant matrix, and the solution to the system Eq.~\eqref{eq:ODES-2} with initial condition~\eqref{eq:initial_condition} exists and is unique. According to Appendix, the solution is:

\begin{equation}
    \label{eq:master_solution}\mathbf{P}(t) = \mathbf{P}(t_{0}) e^{(t-t_0)\mathbf{A}(V_\mathrm{G}, T)}
\end{equation}

Eq.~\eqref{eq:master_solution} describes the probabilities of defect $\alpha$ in each state during the time interval $[t_{0}, t_{0}+ \Delta t]$. It is worth noting that the above method can be applied to defects with multiple charge states (e.g., $M$, $N$, and $K$ configurations for $q=+1$, $0$, and $-1$, respectively), i.e., the capture and emission rate associated with each defect charge state can be considered, enabling RASP to address reliability issues induced by multilevel defects. In this case, when $\frac{d\mathbf{P}(t)}{dt}=0$, the above method naturally reduces to the Sah-Shockley statistics of carrier generation and recombination in semiconductors \cite{Wang2025JACS}.

From Eq.~\eqref{eq:master_solution}, we obtain the charge state of defect $\alpha$ under various device operating conditions. Combining this with Eq.~\eqref{eq:6_discrete}, we can calculate the threshold voltage shift caused by charged defects at different times:
\begin{equation}
    \label{vth}\Delta V_{\mathrm{th}}(t) = V_{\mathrm{traps}}(t) - V_{\mathrm{traps}}(0)
\end{equation}

\section{FRAMEWORK AND MODULES OF RASP}\label{sec:framework}

	    \begin{figure*}[t]
        \centering
        \includegraphics[width=\textwidth]{
            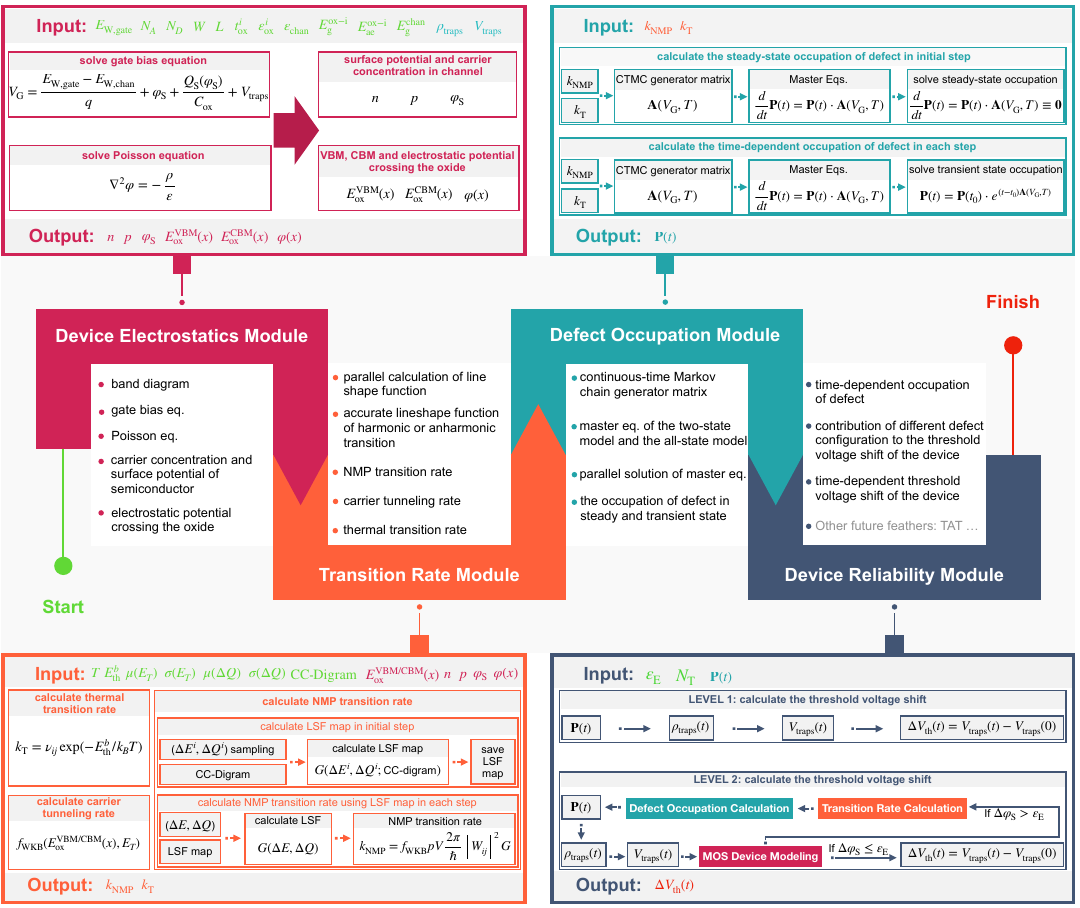
        }
        \caption{\red{Framework of Reliability Ab initio Simulation Package (RASP). RASP consists of four integrated modules: (1) \redc{Device Electrostatics Module} calculates the band diagram, electrostatic potential distribution, and carrier concentration; (2) \redc{Transition Rate Module} computes NMP transition rates, carrier tunneling rates, and thermal transition rates; (3) \redc{Defect Occupation Module} solves the master equations to obtain steady-state and time-dependent probabilities of defect in each state under the competition between all possible transitions.; (4) \redc{Device Reliability Module} evaluates the threshold voltage shift at two schemes (LEVEL 1 or LEVEL 2).}}
        \label{Fig:Framework}
    \end{figure*}


    \begin{table*}[t]
	\centering
	\renewcommand{\arraystretch}{1.35}
	\setlength{\tabcolsep}{8pt}
	\begin{tabular}{|>{\centering\arraybackslash}m{3cm}|>{\arraybackslash}m{8cm}|>{\centering\arraybackslash}m{2cm}|>{\centering\arraybackslash}m{2cm}|}
	\hline
	\text{Parameter Category} & \text{Parameter Name} & \text{Symbol} & \text{Unit} \\ \hline
	\multirow{11}{*}{Device Parameters} 
 	& Channel material & / & / \\ \cline{2-4}
 	& Channel length & $L$ & m \\ \cline{2-4}
 	& Channel width & $W$ & m \\ \cline{2-4}
 	& Channel electron affinity & $E_\text{ae}^\text{channel}$ & eV \\ \cline{2-4}
 	& Channel doping concentration & ${N_A}$/${N_D}$ & m$^{-3}$ \\ \cline{2-4}
 	& Channel bandgap & $E_\mathrm{g}^\mathrm{channel}$ & eV \\ \cline{2-4} 
 	& Oxide material & / & / \\ \cline{2-4}
 	& Oxide thickness & $t_\mathrm{{ox}}$ & nm \\ \cline{2-4}
 	& Oxide electron affinity & $E_\mathrm{ae}^\mathrm{ox}$ & eV \\ \cline{2-4}
 	& Oxide bandgap & $E_\mathrm{g}^\mathrm{ox}$ & eV \\ \cline{2-4}
 	& Gate work function & $E_\mathrm{W, gate}$ & eV \\ \hline
	\multirow{7}{*}{Defect Parameters}
 	& Defect type & acceptor/donor & / \\ \cline{2-4}
 	& Defect concentration & $N_\mathrm{T}$ & m$^{-3}$ \\ \cline{2-4}
 	& Defect distribution range & $x_\mathrm{T}^\mathrm{min}$, $x_\mathrm{T}^\mathrm{max}$ & nm \\ \cline{2-4}
 	& Defect transition level & $E_{T}$ & eV \\ \cline{2-4}
 	& Defect $\Delta Q$ & $\Delta Q$ & amu$^{1/2}\cdot$\AA \\ \cline{2-4}
 	& PES for defect NMP transition pathways & CC diagram & / \\ \cline{2-4}
 	& Energy barriers for defect thermal transition pathways & $E_\text{th}^b$ & eV \\ \cline{2-4}
 	\hline
	\multirow{2}{*}{Stress/Recovery} 
	& Stress voltage & $V_\mathrm{G}^\mathrm{stress}$ & V \\ \cline{2-4}
 	& Recovery voltage & $V_\mathrm{G}^\mathrm{recovery}$ & V \\ \cline{2-4}
	\multirow{1}{*}{Parameters}
 	& Temperature & $T$ & K \\ \hline
	\end{tabular}
	\caption{Device, defect, and stress/recovery parameters required by RASP.}
	\label{tab:input_parameters}
	\end{table*}

The previous sections have discussed the microscopic processes of carrier capture and emission by defects in the gate dielectric layer, as well as their effects on macroscopic device parameters such as threshold voltage. To enable quantitative prediction from atomic-scale defect physics to device-level reliability degradation, we develop Reliability ab initio Simulation Package (RASP). RASP takes defect parameters obtained from first-principles calculations as input and, combined with given device geometry and material parameters, performs accurate simulations of reliability issues caused by each type of defect. The required input parameters are listed in Table~\ref{tab:input_parameters}, where all defect parameters can be obtained from first-principles calculation software such as DASP~\cite{RN20} and doped~\cite{dopedKavanagh2024}. As shown in Fig.~\ref{Fig:Framework}, RASP consists of four modules. First, the \redc{Device Electrostatics Module} solves for the channel surface potential, electrostatic potential in the oxide layer, and device band structure under different operating conditions. Then, the \redc{Transition Rate Module} calculates the rates of relevant transitions during defect carrier capture and emission processes. Subsequently, the \redc{Defect Occupation Module} solves for the time evolution of the probabilities of defect in each state under the competition between all possible transitions. Finally, the \redc{Device Reliability Module} analyzes the quantitative impact of defects on device reliability (such as BTI). In the following, we will give a detailed description of each module.

	\redc{\subsection{Device Electrostatics Module}}
	
	As described in Sec.~\ref{sec:capture_and_emission_rate}, the carrier capture and emission processes by defects in the gate oxide layer are closely related to the device band structure, electrostatic potential distribution, and carrier concentration at the channel. Therefore, accurately characterizing the band structure and electrostatic potential distribution cross the device under a given bias is fundamental for precisely describing the dynamics of defect carrier capture/emission and predicting reliability issues such as threshold voltage shift. To this end, we develop the \redc{Device Electrostatics Module}. Its main function is as follows: given the device geometry (such as channel length $L$, channel width $W$, and oxide layer thicknesses $t_\text{ox}^i$) and material parameters (such as metal gate work function $E_\text{W,gate}$, oxide layer dielectric constants $\varepsilon_\mathrm{ox}^i$, oxide layer bandgaps $E_\text{g}^\text{ox-i}$, oxide layer electron affinities $E_\mathrm{ae}^\text{ox-i}$, channel doping concentrations ${N_A}$ and ${N_D}$, channel bandgap $E_\mathrm{g}^\text{channel}$, and channel electron affinity $E_\text{ae}^\text{channel}$), the module accurately calculates the band structure of the MOS device, the electrostatic potential distribution in the gate oxide layer, and the carrier concentration at the channel under different gate voltages.

The implementation consists of two parts. First, by solving the gate voltage equation (Eq.~\eqref{eq:4}), the module determines the voltage drop across the oxide layer $V_\mathrm{ox}$ and the surface potential at the channel $\varphi_\mathrm{S}$ for different gate voltages $V_\mathrm{G}$. This yields the channel carrier concentrations (\textbf{$n$, $p$}) and the average electric field in the oxide layer under different gate voltages. Second, based on the boundary potentials on both sides of the oxide layer and the spatial distribution of charged defects within the oxide layer, the module solves the Poisson equation to obtain the potential $\varphi(x)$ at any position within the oxide layer and the corresponding oxide band edge positions ($E_\text{ox}^\text{VBM}(x)$, $E_\text{ox}^\text{CBM}(x)$).

The physical quantities calculated by the \redc{Device Electrostatics Module} ($n$, $p$, $\varphi_\mathrm{S}$, $E_\text{ox}^\text{VBM}(x)$, $E_\text{ox}^\text{CBM}(x)$, $\varphi(x)$) are essential for determining carrier tunneling probabilities, calculating position-dependent capture/emission rates, and ultimately accurately predicting the impact on threshold voltage shift.

    \redc{\subsection{Transition Rate Module}}

    During device operation, charged defects in the gate dielectric layer dynamically change their charge states by exchanging carriers with the channel, thereby affecting the electrostatic potential distribution and threshold voltage ($V_{\mathrm{th}}$) of the device. 
    {\red{Taking the process of a single defect capturing carriers from the channel as an example, this process involves two mechanisms: 
(i) \emph{carrier injection} carriers in the channel undergo quantum tunneling across the oxide energy barrier and arrive at the defect site.; and 
(ii) \emph{carrier capture}, carriers are captured by the defect through nonradiative multiphonon and thermal transition.}
    The emission process, in which carriers are released from the defect to the channel, follows the reverse sequence. The \redc{Transition Rate Module} calculates the rates of all transition pathways involved in defect carrier capture and emission.

\red{For carrier injection}, the tunneling behavior can be described using a one-dimensional quantum tunneling model. Under the WKB (Wentzel--Kramers--Brillouin) approximation, the carrier tunneling probability is determined by the defect position ($x_\text{T}$), the potential field between the defect and the channel ($E_\text{ox}^\text{VBM/CBM}(x)$), and the carrier energy ($E$) (see Eq.~\eqref{WKB}). \red{For carrier capture}, due to the inherent low symmetry of amorphous gate dielectrics, certain defects within them can adopt multiple configurations in each charge state. Since these configurations have similar formation energies, all possible transitions among them must be considered when simulating the impact of defects on device reliability. These transitions can be classified into two categories based on whether a charge state change occurs: (i) thermal transitions between configurations with the same charge state; (ii) NMP transitions between different charge states. For thermal transitions, the rate follows transition state theory and can be calculated using Eq.~\eqref{eq:thermal}. For NMP transitions, as described in Sec.~\ref{sec:capture_and_emission_rate}, the rate can be calculated using Eq.~\eqref{eq:H-18}.

Accurate calculation of NMP rates is crucial for simulating carrier capture processes. To balance the speed and accuracy requirements of device/circuit-level simulations, we adopt an LSF-map strategy that involves ``sampling and fitting the LSF (lineshape function) surface first, then interpolating for fast LSF calculation''. During the initialization phase, we sample $(\Delta E^i, \Delta Q^i)$ as described in Sec.~\ref{sec:capture_and_emission_rate} and compute $G(\Delta E^i, \Delta Q^i)$ point by point in parallel. Based on the computed $\{G(\Delta E^i, \Delta Q^i)\}$, we construct an LSF-map ($G(x, y)$, $(x, y) \in \{(\Delta E^i, \Delta Q^i)\}$). During the transition rate calculation at each time step, we simply interpolate from the LSF-map based on the current $(\Delta E, \Delta Q)$ to rapidly obtain $G(\Delta E, \Delta Q)$, which is then substituted into Eq.~\eqref{eq:H-18} to calculate the carrier capture coefficient. This method enables fast and accurate calculation of NMP rates.
	
	\subsection{\redc{Defect Occupation Module}}
	
	As discussed above, the charge state of defects in the gate dielectric layer is no longer dominated by a single mechanism, but is jointly determined by the relative strengths of multiple rate channels under the current bias and temperature conditions. Therefore, after the \redc{Transition Rate Module} computes the rates of various transition mechanisms related to defect carrier capture and emission ($k_\text{NMP}$, $k_\text{T}$), accurately evaluating the charge state of defects under the competition of multiple mechanisms is crucial for quantifying their impact on device performance, especially threshold voltage shift. The \redc{Defect Occupation Module} is designed to solve for the charge states of various defects under a given time-dependent bias voltage. Specifically, when the device remains under steady-state bias conditions (such as the off-state) for an extended period, the defect charge state approaches the steady-state value given by Eq.~\eqref{eq:steady}. During subsequent time-dependent simulations, the applied gate voltage varies with time, causing the rates of various mechanisms to change accordingly, and the defect charge state dynamically adjusts according to Eq.~\eqref{eq:master_solution}.
	
	\subsection{\redc{Device Reliability Module}}
	The \redc{Device Electrostatics Module}, Transition Rate Calculation, and \redc{Defect Occupation Module}s provide a detailed description of the carrier capture and emission processes \redcnew{induced} by defects in the gate dielectric layer, and simulate the time evolution of \redcnew{probabilities of }defects \redcnew{in each state} under time-dependent bias voltages. This enables further quantitative simulation of various reliability issues caused by defects during device operation. Based on these outputs, we develop the \redc{Device Reliability Module}, which evaluates the impact of \redcnew{probabilities of }defects \redcnew{in each state} at different times on critical device parameters (such as threshold voltage, on-state current, and leakage current), thereby enabling quantitative simulation of failure mechanisms such as BTI, RTN, and TAT. In the current version of RASP, this module primarily focuses on quantitative simulation of BTI.

As described in Sec.~\ref{sec:MOS}, there exists an intrinsic coupling between defect charge states and the electrostatic potential distribution of the device. 
\delete{The charge states of defects and the device electrostatic potential are strongly coupled.}On one hand, device electrostatic potential influence defect charge states: the carrier capture and emission rate of defects in the gate oxide layer depend on the device band structure, electrostatic potential distribution, and carrier concentration at the channel. On the other hand, defect charge states influence device electrostatic potential: as a type of space charge, the evolution of the probabilities of defect in charged state also affects the electrostatic potential distribution across the device.
Based on the strength of this interaction, we introduce two different simulation schemes in the \redc{Device Reliability Module}: LEVEL 1 and LEVEL 2, corresponding to weak coupling and strong coupling scenarios, respectively.

LEVEL 1 is suitable for cases where the defect concentration in the device is relatively low. In this case, the change of the electrostatic potential \delete{caused }\redcnew{induced} by charged defects is small \redcnew{and can be neglected}. The impact on device electrical characteristics can be analyzed from two \delete{aspects }\redcnew{perspectives}: the gate voltage equation and the Poisson equation.
\begin{itemize}
    \item Gate voltage equation: Carrier capture and emission induced by defects cause changes in the trapped charge $Q_{\mathrm{traps}}$ in the oxide layer, which in turn affects the $V_{\mathrm{traps}}$ term in the gate voltage equation, leading to a drift in the channel surface potential $\varphi_\mathrm{S}$. This is the direct cause of device threshold voltage shift ($\Delta V_{\mathrm{th}}$).
    \item Poisson equation: Due to the low defect concentration, the change in space charge density caused by carrier capture and emission is limited, and the impact on the potential distribution inside the oxide layer can be neglected. Therefore, the potential in the oxide layer can be approximated as maintaining a linear distribution, and the electric field remains uniform.
\end{itemize}

In this scheme, the calculation procedure is simplified (the corresponding pseudocode for the LEVEL 1 scheme is presented in Algorithm~\ref{alg:level1}):
\begin{itemize}
    \item Based on the defect charge state $\mathbf{P}(t)$ and its spatial distribution output by the \redc{Defect Occupation Module}, the contribution of charged defects to the total voltage drop $V_{\mathrm{traps}}(t)$ at different times is directly calculated, yielding the threshold voltage shift using Eq.~\eqref{vth}.
\end{itemize}

\begin{algorithm}[h]
\SetAlgoLined
\caption{Calculation of $\Delta V_\text{th}$ at time $t$ using LEVEL 1 scheme in RASP}
\label{alg:level1}
\KwInput{$\mathbf{P}(t)$, $\rho_{\mathrm{traps}}(x,t)$, $V_\text{traps}(0)$}
\KwOutput{$\Delta V_\text{th}(t)$}

\tcp{Assumption: Low defect concentration, negligible perturbation to oxide electrostatic potential}


\quad $V_{\mathrm{traps}}(t) \gets \mathbf{P}(t), \rho_{\mathrm{traps}}(x,t)$\;

\quad $\Delta V_\text{th}(t) \gets V_{\mathrm{traps}}(t) - V_{\mathrm{traps}}(0)$\;

\Return{$\Delta V_\mathrm{th}(t)$}
\end{algorithm}

This method does not require solving the Poisson equation at each time step, resulting in high computational efficiency. It is suitable for rapid evaluation of BTI degradation in devices with low defect density.

LEVEL 2 is designed for cases with high defect concentration. In this case, carrier capture and emission not only affect the gate voltage equation but also cause significant non-uniform perturbations to the electrostatic potential distribution inside the oxide layer. The dynamic processes of carrier capture and emission \redcnew{induced} by defects must be considered at both the gate voltage equation and Poisson equation levels:
\begin{itemize}
    \item Gate voltage equation: Same as LEVEL 1, changes in the defect charge state $\mathbf{P}(t)$ modify the channel surface potential $\varphi_\mathrm{S}$ and threshold voltage $V_{\mathrm{th}}$ through the $V_{\mathrm{traps}}$ term.
    \item Poisson equation: Due to the high defect concentration, carrier capture and emission by a large number of defects lead to non-negligible space charge change inside the oxide layer. This causes the electrostatic potential distribution inside the oxide layer to deviate significantly from a linear relationship, and the electric field is no longer uniform. This nonlinear potential perturbation changes the band edge energy of the oxide at the defect location, thereby affecting the carrier capture and emission rates of the corresponding defects, and consequently influencing the defect charge states.
\end{itemize}


As shown in Fig.~\ref{Fig:Framework}, for each time $t$, the defect charge states and the potential distribution in the gate dielectric layer must be solved through self-consistent iteration. The specific procedure is as follows (where $i$ denotes the iteration step), and the corresponding pseudocode for the LEVEL 2 scheme is presented in Algorithm~\ref{alg:self_consistent}:
\begin{itemize}
    \item Based on the defect charge state ${\mathbf{P}(t)}^{i=0}$ and its spatial distribution output by the \redc{Defect Occupation Module}, calculate ${V_{\mathrm{traps}}(t)}^{i=0}$ and solve the gate voltage equation to obtain ${\varphi_\text{S}(t)}^{i=1}$. Substitute ${\rho_{\mathrm{traps}}(x)}^{i=0}$ into the Poisson equation to solve for ${\varphi_\text{ox}(x,t)}^{i=1}$, ${E_\text{ox}^\text{VBM}(x,t)}^{i=1}$, and ${E_\text{ox}^\text{CBM}(x,t)}^{i=1}$.
    \item If $\left|{\varphi_\text{S}(t)}^{i=1}-{\varphi_\text{S}(t)}^{i=0}\right|\leq\varepsilon_\text{E}$, where $\varepsilon_\text{E}$ is a user-defined tolerance threshold, the defect charge states and device potential have reached equilibrium. The threshold voltage shift caused by defects at this time is $\Delta V_\text{th}(t) = {V_{\mathrm{traps}}(t)}^{i=0}-V_{\mathrm{traps}}(t=0)$. If $\left|{\varphi_\text{S}(t)}^{i=1}-{\varphi_\text{S}(t)}^{i=0}\right|>\varepsilon_\text{E}$, the defect charges and device potential have not reached self-consistent equilibrium, and ${k_\text{NMP}(t)}^{i=1}$ and ${k_\text{T}}^{i=1}$ need to be recalculated based on ${E_\text{ox}^\text{VBM}(x,t)}^{i=1}$, ${E_\text{ox}^\text{CBM}(x,t)}^{i=1}$, and ${\varphi_\text{S}(t)}^{i=1}$.
    \item Solve for the defect charge state ${\mathbf{P}(t)}^{i=1}$ based on the recalculated ${k_\text{NMP}(t)}^{i=1}$ and ${k_\text{T}}^{i=1}$.
    \item Iterate until $\left|{\varphi_\text{S}(t)}^{i+1}-{\varphi_\text{S}(t)}^{i}\right|\leq\varepsilon_\text{E}$.
\end{itemize}

\begin{algorithm}[h]
\SetAlgoLined
\caption{Calculation of $\Delta V_\text{th}$ at time $t$ using LEVEL 2 scheme in RASP}
\label{alg:self_consistent}

\KwInput{$\mathbf{P}^0$, $\rho_{\text{traps}}^0$, $\varepsilon_\text{E}$, $\varphi_\text{S}^0$, $V_\text{traps}(0)$}
\KwOutput{$\Delta V_\text{th}(t)$}

$i \gets 0$\;
$\text{converged} \gets \text{False}$\;

\While{not converged}{
    $V_{\mathrm{traps}}^i(t) \gets \mathbf{P}^i, \rho_{\mathrm{traps}}^i$\;
    
    Solve gate voltage equation (Eq.~\eqref{eq:4})\;
    \quad $\rightarrow \varphi_\text{S}^{i+1}$\;

    \If{$|\varphi_\mathrm{S}^{i+1} - \varphi_\mathrm{S}^i| \leq \varepsilon_\text{E}$}{
        $\Delta V_\text{th}(t) \gets V_{\mathrm{traps}}^i(t) - V_{\mathrm{traps}}$(0)\;
        $\text{converged} \gets \text{True}$\;
        \Return{$\Delta V_\mathrm{th}(t)$}
    }
    \Else{
        Solve Poisson equation (Eq.~\eqref{eq:possion})\;
    \quad $\rightarrow \varphi_\mathrm{ox}(x)^{i+1}, E_\mathrm{ox}^{\mathrm{VBM}}(x)^{i+1}, E_\mathrm{ox}^{\mathrm{CBM}}(x)^{i+1}$\;
        Compute $k_{\mathrm{NMP}}^{i+1}$ and $k_\text{T}^{i+1}$\;
        Solve master equation for $\mathbf{P}^{i+1}$\;
        Update $\rho_{\mathrm{traps}}^{i+1}$\;
        $i \gets i+1$\;
    }
}

\end{algorithm}

Through this approach, the LEVEL 2 scheme can accurately capture the strong coupling \delete{effects }between defect charge states and \redcnew{electrostatic potential distribution of }device\delete{ potential}, thereby providing more accurate quantitative simulations for reliability issues under high defect concentrations.


	\section{Simulation of NBTI induced by $\mathrm{V_O}$}
	\subsection{$\Delta V_\mathrm{{th}}$ induced by $\mathrm{V_O}$ with only ground-state configuration\redc{s} (Two-state model) }
	
	\begin{figure*}[t]
		\centering
         \includegraphics[width=0.95\textwidth]{
             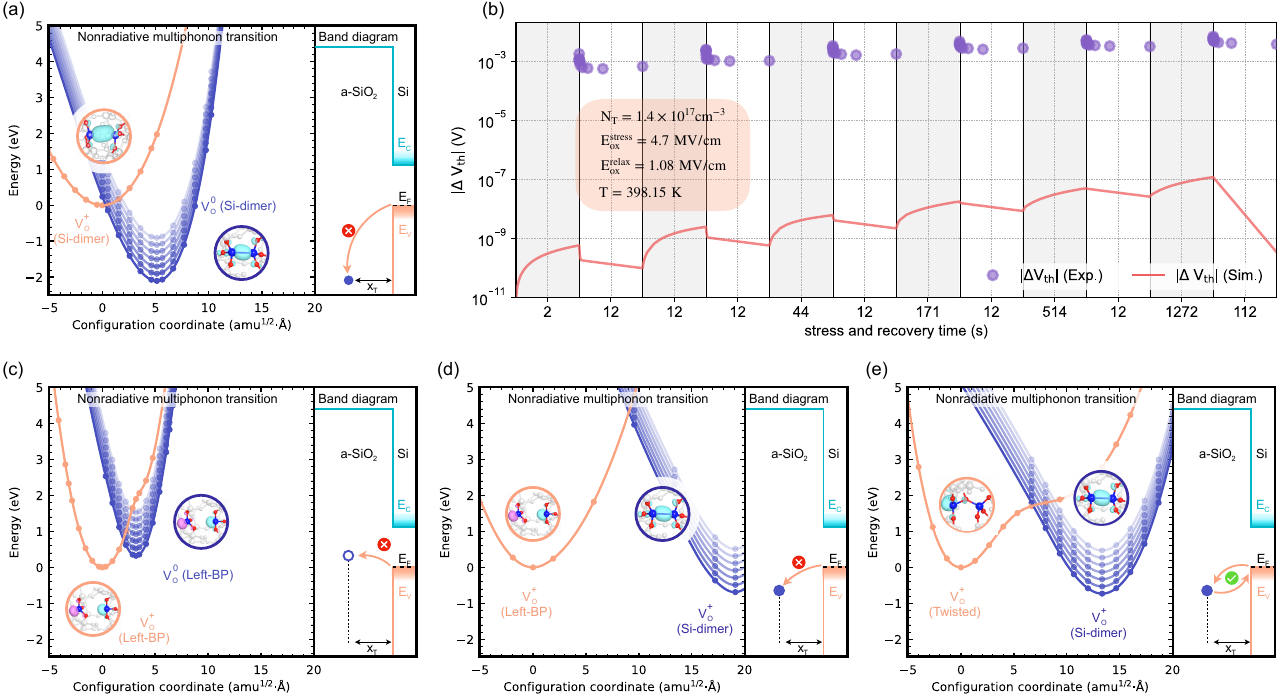
         }
        
         \caption{Impact of trap level, $\Delta Q$, and configuration coordinate diagram on $\Delta V_{\mathrm{th}}$. (a) Configuration coordinate diagram and band diagram for the V$_\mathrm{O}^{0}$ (Si-dimer) $\rightarrow$ V$_\mathrm{O}^{+}$ (Si-dimer) transition: a deep trap level $E_{T}$ results in a large hole-capture barrier. (b) $|\Delta V_{\mathrm{th}}|$ during stress and recovery cycles: experimental data~\cite{Comphy_Rzepa_201849, Comphy3_Dominic_2023} (purple circles) compared with RASP simulation (solid red line). (c) Schematic illustrating an excessively high $E_{T}$: the defect remains in a pre-charged state and is inactive for carrier capture/emission. (d) Nearly harmonic potential energy surface with large $\Delta Q$: the high NMP barrier suppresses carrier capture/emission. (e) Strongly anharmonic potential energy surface involving bond reconfiguration (e.g., twisted configuration): the reduced barrier enables carrier capture/emission and contributes to $\Delta V_{\mathrm{th}}$.}
         \label{Fig:DefectParametersImpact}

	\end{figure*}
	As described in Sec.~\ref{sec:framework}, the all-state model is implemented in RASP. When the all-state model considers only the ground-state configurations before and after carrier capture (emission), it reduces to the two-state model (the proof is given in Appendix). Due to the simplicity of the two-state model, it is still widely used in many studies~\cite{TSM_Case1, TSM_Case2, TSM_Case3}, and therefore RASP also supports the two-state model. In this case, only NMP transitions participate in the defect carrier capture/emission process. As described in Sec.~\ref{sec:capture_and_emission_rate}, the parameters affecting the NMP rate are the defect energy level $E_{T}$, the lattice relaxation $\Delta Q$, and the potential energy surface corresponding to the NMP transition.

	\red{For the defect energy level $E_{T}$, its position relative to the Si valence band maximum (Si-VBM) significantly affects the rate of carrier capture and emission \redcnew{induced} by defects. For example, for V$_{\mathrm{O}}$ formed at 35.4\% of the oxygen sites in a-SiO$_{\mathrm{2}}$, the ground-state configurations of both V$_{\mathrm{O}}^{0}$ and V$_{\mathrm{O}}^{+}$ are the Si-dimer configuration~\cite{guo2024all}. If only the transition between the ground states of neutral and +1 charged V$_{\mathrm{O}}$ is considered (two-state model), only NMP transitions affect $\Delta V_\text{th}$. The CC diagram and the position of the defect energy level relative to Si-VBM for the NMP transition V$_{\mathrm{O}}^{0}$(Si-dimer) $\rightarrow$ V$_{\mathrm{O}}^{+}$(Si-dimer) are shown in Fig.~\ref{Fig:DefectParametersImpact}(a). Since the defect energy level is low (2.1~eV below Si-VBM), the barrier for the defect to capture hole carriers from the channel is high, making the capture process difficult to occur. To quantitatively evaluate its impact on device $\Delta V_\text{th}$, we used RASP to simulate the resulting $\Delta V_\text{th}$, with simulation parameters listed in Table~I in Supplementary Materials. The simulation results are shown in Fig.~\ref{Fig:DefectParametersImpact}(b). It can be seen that the resulting $\Delta V_\text{th}$ is negligible, which is consistent with conclusions from previous studies~\cite{RN64}. On the other hand, if the position of $E_{T}$ relative to Si-VBM is too high, for example, during the recovery phase of a pMOS device, $E_{T}$ remains above the Fermi level $E_F$ (for pMOS, $E_F$ is located near Si-VBM during operation), as shown in Fig.~\ref{Fig:DefectParametersImpact}(c). In this case, the defect tends to be already in a charged state before stress is applied. Defects in this charged state can neither capture new hole carriers under stress nor easily emit hole carriers during the subsequent recovery phase, and therefore do not affect $\Delta V_\text{th}$. Consequently, only defects with energy levels falling within a specific window near Si-VBM can capture/emit carriers from the channel, making them potential sources of device reliability degradation. }
    
    \red{For the lattice relaxation $\Delta Q$, when a defect captures or emits a carrier (such as a hole), the atomic structure around it undergoes relaxation, and $\Delta Q$ is the parameter that quantifies the extent of this relaxation. The effect of $\Delta Q$ on the barrier mainly depends on the specific shape of the potential energy surface (PES) during the NMP transition, which is closely related to the change in defect microscopic configuration during the NMP transition process. If the defect PES exhibits strong harmonic characteristics, as shown in Fig.~\ref{Fig:DefectParametersImpact}(d), a large $\Delta Q$ will directly result in a high barrier for defect hole carrier capture/emission. Even if $E_{T}$ is near Si-VBM, the excessively high barrier will hinder carrier capture/emission, making it difficult to contribute to $\Delta V_\text{th}$. However, if the NMP transition involves bond breaking and reformation (such as V$_{\mathrm{O}}^{0}$(Si-dimer) $\rightarrow$ V$_{\mathrm{O}}^{+}$(twisted)), the corresponding PES will exhibit significant anharmonic characteristics. As shown in Fig.~\ref{Fig:DefectParametersImpact}(e), this anharmonicity of the PES greatly reduces the carrier capture/emission barrier, allowing the capture/emission process to occur and potentially affecting $\Delta V_\text{th}$. Therefore, when simulating the impact of defect-induced NMP processes on device reliability, we need to accurately calculate the defect energy level $E_{T}$, the lattice relaxation $\Delta Q$, and the shape of the PES corresponding to the NMP transition.}

\redc{\subsection{$\Delta V_\mathrm{{th}}$ induced by $\mathrm{V_O}$ with all possible configurations on one O site}}
	So far, we have analyzed the effects of defect parameters on the NMP transition process and simulated their impact on device threshold voltage shift ($\Delta V_\text{th}$) with the two-state model (considering only NMP transitions between the ground-state configurations of neutral and +1 charged V$_{\mathrm{O}}$). \red{However, the two-state model is insufficient to explain two typical types of defect behavior observed in experiments~\cite{RN69,RN64}. To describe these complex dynamics of defects, a series of models have been proposed in recent years, such as the Harry-Diamond-Labs (HDL) switching trap model~\cite{HDL} and the four-state model~\cite{RN54}. The key improvement of these models lies in the introduction of metastable configurations of V$_{\mathrm{O}}$ and the transitions among these configurations.}

		\begin{figure}[htbp!]
		\centering
         \includegraphics[width=0.49\textwidth]{
             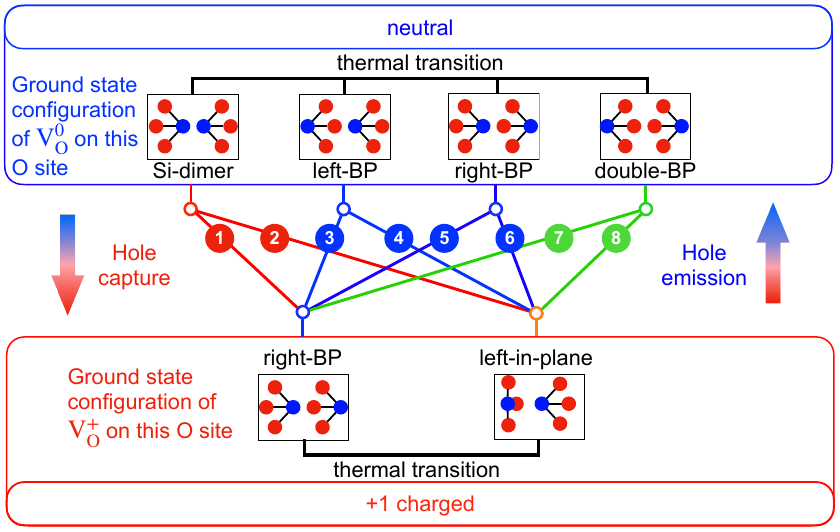
         }
         \caption{\red{Stable and metastable configurations of V$_\mathrm{O}^{0}$ and V$_\mathrm{O}^{+}$ formed at the selected O site, and the corresponding transition pathways. 
In the neutral charge state, V$_\mathrm{O}^{0}$ exists four configurations, i.e., Si-dimer, left-back-projected (left-BP), right-back-projected (right-BP), and double-back-projected (double-BP), where the Si-dimer configuration is the ground-state. In the $+1$ charge state, V$_\mathrm{O}^{+}$ exhibits two configurations, i.e., left-in-plane and right-back-projected (right-BP), where the left-in-plane configuration is the ground-state. 
Thermal transitions occur between configurations with the same charge state and are denoted by the horizontal black lines. NMP transitions (hole capture/emission) occur between configurations of different charge states and are denoted by the red, blue, and green lines.}}
         \label{Fig:Structure_046}

	\end{figure}
	
	\begin{figure}[t]
		\centering
         \includegraphics[width=0.45\textwidth]{
             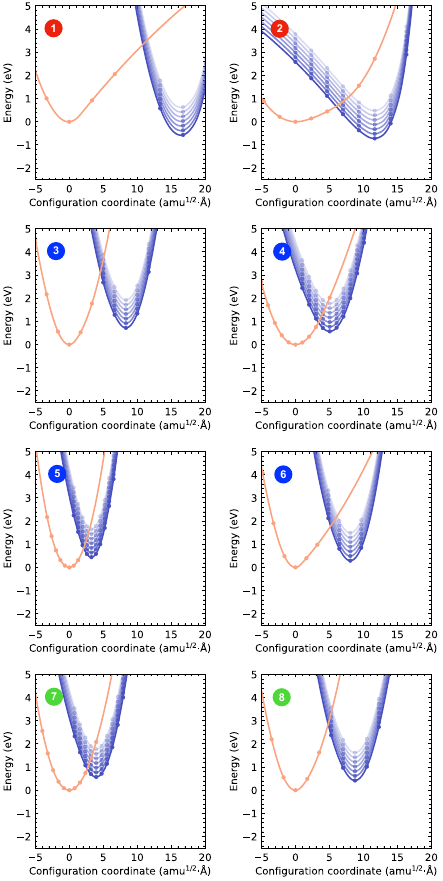
         }
         
         \caption{\red{Configuration coordinate diagrams for the 8 NMP transition pathways shown in Fig. \ref{Fig:Structure_046}. The numbers correspond to those in Fig. \ref{Fig:Structure_046}. In each panel, the orange curve represents the potential energy surface (PES) of +1 charged V$_{\mathrm{O}}$, and the blue curves represent the PES of neutral V$_{\mathrm{O}}$. Under negative gate bias (stress), the PES of neutral V$_{\mathrm{O}}$ shifts upward, as indicated by the multiple blue curves in each panel.}}
         \label{Fig:CCD_046}

	\end{figure}	
	We take V$_{\mathrm{O}}$ formed at a specific oxygen site in a-SiO$_{\mathrm{2}}$ as an example to analyze in detail the impact of considering metastable configurations of neutral and +1 charged V$_{\mathrm{O}}$ on $\Delta V_\text{th}$ and $\tau_e$. As shown in Fig.~\ref{Fig:Structure_046}, V$_{\mathrm{O}}^{0}$ formed at this oxygen site has four possible configurations, among which Si-dimer is the ground-state configuration. V$_{\mathrm{O}}^{+}$ formed at this site has two possible configurations, among which right-BP is the ground-state configuration and left-in-plane is the metastable configuration. Notably, the left-in-plane configuration is unique to V$_{\mathrm{O}}^{+}$ in a-SiO$_{\mathrm{2}}$. For V$_{\mathrm{O}}$ formed at this oxygen site, the carrier capture/emission process involves 8 NMP transition pathways. If only the NMP transition between the ground states of V$_{\mathrm{O}}^{0}$ and V$_{\mathrm{O}}^{+}$ is considered, as in the two-state model (see pathway 1 in Fig.~\ref{Fig:Structure_046}), the CC diagram corresponding to pathway 1 in Fig.~\ref{Fig:CCD_046} shows that the carrier capture/emission barriers for this process are both large. Under the two-state model, V$_{\mathrm{O}}$ formed at this oxygen site can hardly undergo carrier capture/emission. We simulated the impact of this process on device threshold voltage using the parameters in Table~I in Supplementary Materials. As shown in the upper panel of Fig.~\ref{Fig:Two_and_All_state_model}, this process has almost no effect on device threshold voltage. What if we adopt the all-state model and consider all defect configurations and all transitions among them? As shown in the lower panel of Fig.~\ref{Fig:Two_and_All_state_model}, the carrier capture/emission by V$_{\mathrm{O}}$ formed at this site causes non-negligible threshold voltage shift. This indicates that the metastable configurations of V$_{\mathrm{O}}^{0}$ and V$_{\mathrm{O}}^{+}$ not only participate in the carrier capture/emission process of V$_{\mathrm{O}}$ but also play an important role.

Next, we quantitatively analyze how metastable configurations participate in the carrier capture/emission process. As shown in Fig.~\ref{Fig:CCD_046} (\redcnew{pathway 1}), since the energy level between the ground states of V$_{\mathrm{O}}^{0}$ and V$_{\mathrm{O}}^{+}$ lies below $E_F$, V$_{\mathrm{O}}$ is neutral at steady state. For the stress process, although V$_{\mathrm{O}}^{0}$(Si-dimer) can hardly capture a hole to transform into V$_{\mathrm{O}}^{+}$(right-BP), \delete{for pathway 2, }V$_{\mathrm{O}}^{0}$(Si-dimer) can relatively easily capture a hole under stress conditions to transform into the metastable V$_{\mathrm{O}}^{+}$(left-in-plane) \redcnew{via pathway 2}, which then overcomes a thermal barrier (0.87~eV) to transform into V$_{\mathrm{O}}^{+}$(right-BP). For the recovery process, as shown in the CC diagram of pathway 1 in Fig.~\ref{Fig:CCD_046}, the ground-state configuration V$_{\mathrm{O}}^{+}$(right-BP) also has difficulty \delete{releasing}\redcnew{emitting} a hole to transform into V$_{\mathrm{O}}^{0}$(Si-dimer). However, it can first overcome a thermal barrier (0.99~eV) to transform into V$_{\mathrm{O}}^{+}$(left-in-plane), which can then relatively easily emit a hole to transform into V$_{\mathrm{O}}^{0}$(Si-dimer) (as shown in the CC diagram of pathway 2 in Fig.~\ref{Fig:CCD_046}). Furthermore, as shown in the CC diagrams of pathways 5 and 7 in Fig.~\ref{Fig:CCD_046}, the barriers for V$_{\mathrm{O}}^{+}$(right-BP) to emit a hole and transform into V$_{\mathrm{O}}^{0}$(right-BP) or V$_{\mathrm{O}}^{0}$(double-BP) are small. Subsequently, V$_{\mathrm{O}}^{0}$(right-BP) and V$_{\mathrm{O}}^{0}$(double-BP) can overcome thermal barriers to transform into V$_{\mathrm{O}}^{0}$(Si-dimer). None of these processes are considered in the two-state model, which consequently underestimates the contribution of this V$_{\mathrm{O}}$ to device threshold voltage shift.

	\begin{figure}[t]
		\centering
         \includegraphics[width=0.50\textwidth]{
             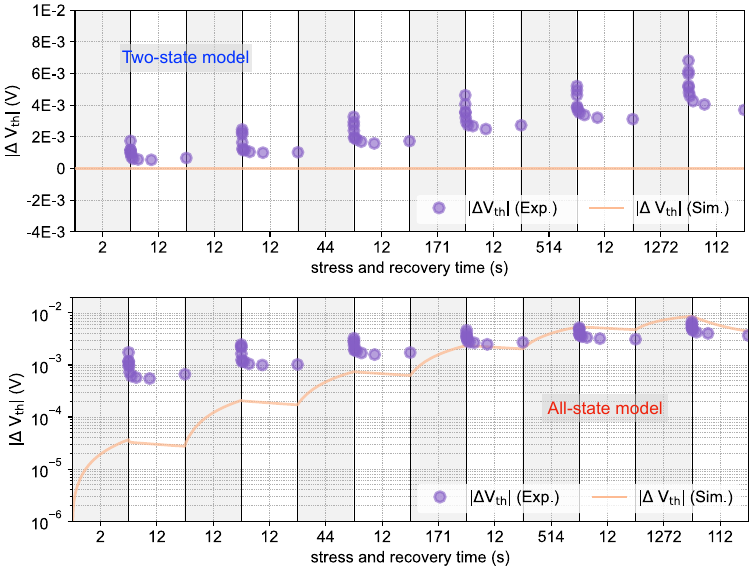
         }
         \caption{\red{Comparison of threshold voltage shift ($\Delta V_{\mathrm{th}}$) simulated by RASP using the two-state model and the all-state model during stress and recovery cycles. The upper panel shows the two-state model, which includes only the NMP transition between the ground-state configurations, i.e., V$_{\mathrm{O}}^{0}$ (Si-dimer) $\leftrightarrow$ V$_{\mathrm{O}}^{+}$ (right-BP), yielding a negligible $\Delta V_{\mathrm{th}}$. The lower panel shows the all-state model, which accounts for all possible defect configurations as shown in Fig. \ref{Fig:Structure_046} and all NMP/thermal transitions among them. The simulated $\Delta V_{\mathrm{th}}$ shows that the V$_\text{O}$ formed on this O site play a non-negligible role in NBTI. Orange lines denote simulation results of RASP, and purple circles denote experimental data~\cite{Comphy_Rzepa_201849, Comphy3_Dominic_2023}.}}
         \label{Fig:Two_and_All_state_model}

	\end{figure}

	\begin{figure}[t]
		\centering
         \includegraphics[width=0.49\textwidth]{
             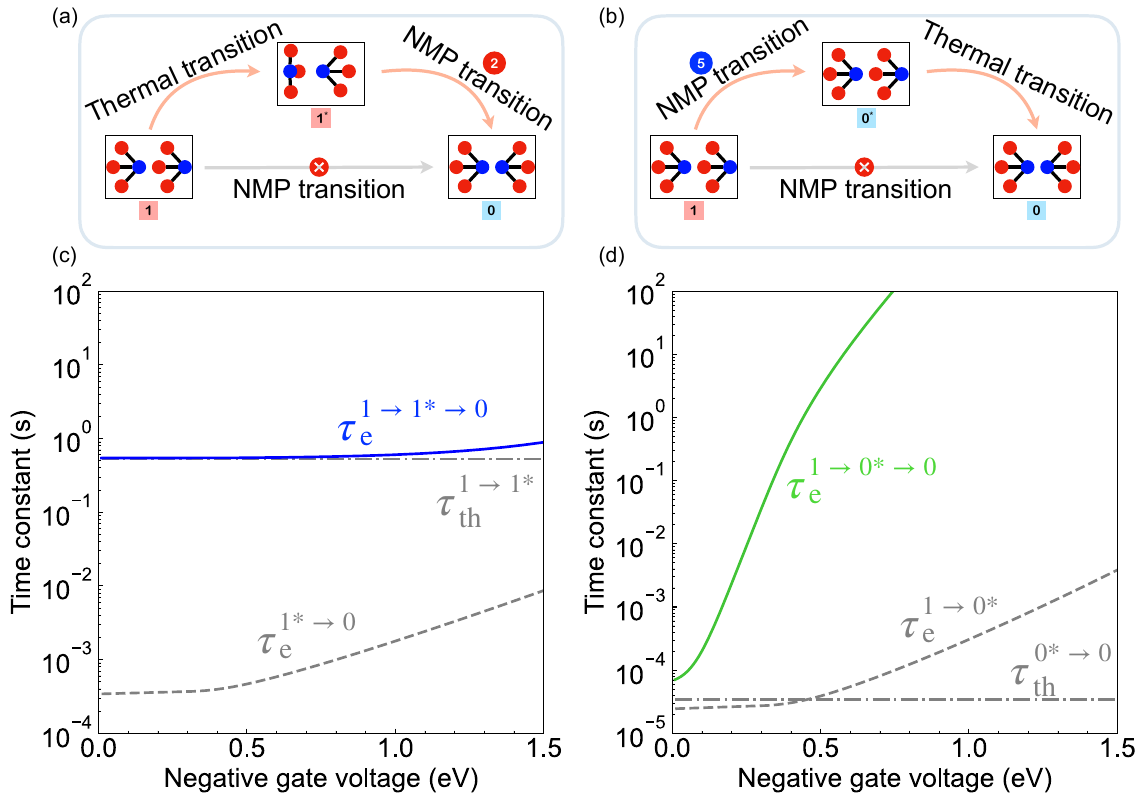
         }

\caption{\red{Role of metastable configurations in the emission kinetics of V$_\mathrm{O}$. 
(a) Emission pathway from the V$_\mathrm{O}^{+}$ ground-state configuration (right-back-projected, labeled as $1$) to the V$_\mathrm{O}^{0}$ ground-state configuration (Si-dimer, labeled as $0$) via the V$_\mathrm{O}^{+}$ metastable configuration (left-in-plane, labeled as $1^{*}$). This pathway consists of a thermal transition ($1\!\rightarrow\!1^{*}$) followed by an NMP transition ($1^{*}\!\rightarrow\!0$). 
(b) Emission pathway from $1$ to $0$ via the V$_\mathrm{O}^{0}$ metastable configuration (right-BP, labeled as $0^{*}$). This pathway consists of an NMP transition ($1\!\rightarrow\!0^{*}$) followed by a thermal transition ($0^{*}\!\rightarrow\!0$). 
(c) Emission time constant $\tau_{e}$ for the indirect pathway $1\!\rightarrow\!1^{*}\!\rightarrow\!0$ ($\tau_{e}^{1\rightarrow1^{*}\rightarrow 0}$, solid blue) as a function of \redcnew{negative }gate voltage \redcnew{(-V$_\mathrm{G}$)}. Gray dashed lines show the emission time constants of the two constituent steps: thermal transition $1\!\rightarrow\!1^{*}$ ($\tau_{\mathrm{th}}^{1\rightarrow 1^{*}}$) and NMP transition $1^{*}\!\rightarrow\!0$ ($\tau_{e}^{1^{*}\rightarrow 0}$). 
(d) Emission time constant $\tau_{e}$ for the indirect pathway $1\!\rightarrow\!0^{*}\!\rightarrow\!0$ ($\tau_{e}^{1\rightarrow0^{*}\rightarrow 0}$, solid green) as a function of \redcnew{negative }gate voltage \redcnew{(-V$_\mathrm{G}$)}. Gray dashed lines show the emission time constants of the two constituent steps: NMP transition $1\!\rightarrow\!0^{*}$ ($\tau_{e}^{1\rightarrow 0^{*}}$) and thermal transition $0^{*}\!\rightarrow\!0$ ($\tau_{\mathrm{th}}^{0^{*}\rightarrow 0}$).}}

         \label{Fig:Fxied_and_Switching_Traps}

	\end{figure}	
	On the other hand, the different dependencies of $\tau_e$ on $V_\mathrm{G}$ for fixed traps and switching traps observed in TDDS experiments also confirm the importance of considering metastable configurations. In the following, we will analyze the effects of neutral and +1 charged metastable configurations on the emission time constant during the recovery process, respectively.

For transition processes involving the +1 charged V$_{\mathrm{O}}$ metastable configuration, taking the path $1\left[\mathrm{V_O^+}\left(\text{right-BP}\right)\right]$ $\rightarrow$ $1^*\left[\mathrm{V_O^+}\left(\text{left-in-plane}\right)\right]$  $\rightarrow$ $0\left[\mathrm{V_O^0}\left(\text{Si-dimer}\right)\right]$ as an example, the first passage time for this process can be expressed as:
\begin{equation}
    \begin{aligned}
        \tau_e^{1\rightarrow 1^* \rightarrow 0} &= \tau_\mathrm{th}^{1\rightarrow 1^*} + \tau_{e}^{1^*\rightarrow0}\left(1+\frac{\tau_\mathrm{th}^{1\rightarrow 1^*}}{\tau_\mathrm{th}^{1^*\rightarrow1}}\right) \\
        &=\tau_\mathrm{th}^{1\rightarrow 1^*} + \tau_{e}^{1^*\rightarrow0}\left[1+\exp\left(\frac{E(1^*)-E(1)}{k_B T}\right) \right]
    \end{aligned}
\end{equation}
As shown in the CC diagram of NMP transition 2 in Fig.~\ref{Fig:CCD_046}, when $V_\mathrm{G}$ is small, the corresponding NMP barrier is low, and the thermal transition $1\rightarrow1^*$ becomes the rate-limiting process. Since $\tau_\mathrm{th}^{1\rightarrow 1^*}$ does not vary with $V_\mathrm{G}$, $\tau_e^{1\rightarrow 1^* \rightarrow 0}$ remains nearly constant with $V_\mathrm{G}$, exhibiting the characteristics of fixed traps (as shown in Fig.~\ref{Fig:Fxied_and_Switching_Traps}(a)). As $V_\mathrm{G}$ increases, the NMP barrier increases, and NMP transition 2 becomes the rate-limiting process. Consequently, $\tau_e^{1\rightarrow 1^* \rightarrow 0}$ increases with increasing $V_\mathrm{G}$ at larger $V_\mathrm{G}$ values. 

For transition processes involving the neutral V$_{\mathrm{O}}$ metastable configuration, taking the path $1\left[\mathrm{V_O^+}\left(\text{right-BP}\right)\right]$ $\rightarrow$ $0^*\left[\mathrm{V_O^0}\left(\text{right-BP}\right)\right]$ $\rightarrow$ $0\left[\mathrm{V_O^0}\left(\text{Si-dimer}\right)\right]$ as an example, the first passage time for this process can be expressed as:
\begin{equation}
    \begin{aligned}\label{eq:switching}
        \tau_e^{1\rightarrow 0^* \rightarrow 0} &= \tau_{e}^{1\rightarrow0^*} + \tau_\mathrm{th}^{0^*\rightarrow 0}\left(1+\frac{\tau_{e}^{1\rightarrow0^*}}{\tau_{e}^{0^*\rightarrow1}}\right)\\
        &=\tau_{e}^{1\rightarrow0^*} + \tau_\mathrm{th}^{0^*\rightarrow 0}\left(1+\exp\left[\frac{\left(E_T-E_F\right)}{k_BT}\right]\right)
    \end{aligned}
\end{equation}
From the above equation, $\tau_e^{1\rightarrow 0^* \rightarrow 0}$ depends exponentially on $E_F$. When $V_\mathrm{G}$ is near the device threshold voltage, $E_F$ changes significantly. Therefore, when the recovery voltage $|V_\mathrm{G}|$ decreases to near or below the threshold voltage $|V_\mathrm{th}|$, $\tau_e$ exhibits a sharp nonlinear decrease, as shown in Fig.~\ref{Fig:Fxied_and_Switching_Traps}(b). The process $1\rightarrow 0^* \rightarrow 0$ involving the neutral metastable configuration exhibits typical switching trap characteristics near the threshold voltage ($V_\text{th}$=-0.47 eV).

Therefore, when simulating the impact of defects in the dielectric layer on device reliability, it is necessary not only to consider the ground-state configurations of defects and the NMP transitions between them as in the two-state model, but also to include all possible metastable configurations of defects along with their associated NMP transitions and thermal transitions. All these possible defect configurations and all possible NMP and thermal transitions among them are considered in the all-state model, enabling accurate simulation of the impact of defects in amorphous gate dielectrics on device reliability in RASP.
	
\
	
	\redc{\subsection{$\Delta V_\mathrm{{th}}$ induced by $\mathrm{V_O}$ with all possible configurations on different O sites}}

	Due to the low symmetry of amorphous materials, each atom resides in a different local \redcnew{atomic} environment. 
    The configurations in which V$_{\mathrm{O}}$ can stably exist vary among different oxygen sites in amorphous materials. For example, at some oxygen sites, as shown in Fig.~\ref{Fig:Structure_046}, +1 charged V$_{\mathrm{O}}$ has only two stable configurations: left-in-plane and right-BP. At other oxygen sites, as shown in Fig.~\ref{Fig:QQP_Structure}, left-in-plane, right-in-plane, and right-BP configurations may coexist, and the ground-state structures also differ among different oxygen sites. This means that the ground-state and metastable configurations of V$_{\mathrm{O}}$ in $\alpha$-SiO$_{\mathrm{2}}$ cannot be used to describe defects formed at all oxygen sites in a-SiO$_{\mathrm{2}}$. 
    In RASP, to capture the diversity of defect configurations and energies at different oxygen sites in amorphous materials, the contribution of V$_{\mathrm{O}}$ formed at each oxygen site to $\Delta V_{\mathrm{th}}$ is considered individually.
	
	To illustrate the \delete{effects}\redcnew{impacts} of the low symmetry of amorphous materials on defect parameters and device reliability, we selected ten oxygen sites in a-SiO$_{\mathrm{2}}$, where V$_{\mathrm{O}}$ configurations are the same as those shown in Fig.~\ref{Fig:QQP_Structure}, and analyzed the impact of V$_{\mathrm{O}}$ formed at these oxygen sites on device threshold voltage shift. As shown in Fig.~\ref{Fig:QQP_Structure}, V$_{\mathrm{O}}^{0}$ formed at this type of oxygen site has four possible configurations: Si-dimer, left-back-projected, right-back-projected, and double-back-projected, among which Si-dimer is the ground-state configuration (see Table~II in Supplementary Materials). V$_{\mathrm{O}}^{+}$ formed at these sites has three possible configurations: left-in-plane, right-in-plane, and right-back-projected, among which in-plane (left-in-plane or right-in-plane) is the ground-state configuration (see Table~II in Supplementary Materials). Therefore, there are 12 possible NMP transition pathways between V$_{\mathrm{O}}^{0}$ and V$_{\mathrm{O}}^{+}$ formed at each oxygen site. The transition levels and $\Delta Q$ values corresponding to the 12 NMP pathways at the 10 oxygen sites considered are listed in Table~II and Table~III in Supplementary Materials, respectively.

\begin{figure*}[!htbp]
        \centering
        \includegraphics[width=0.85\textwidth]{
            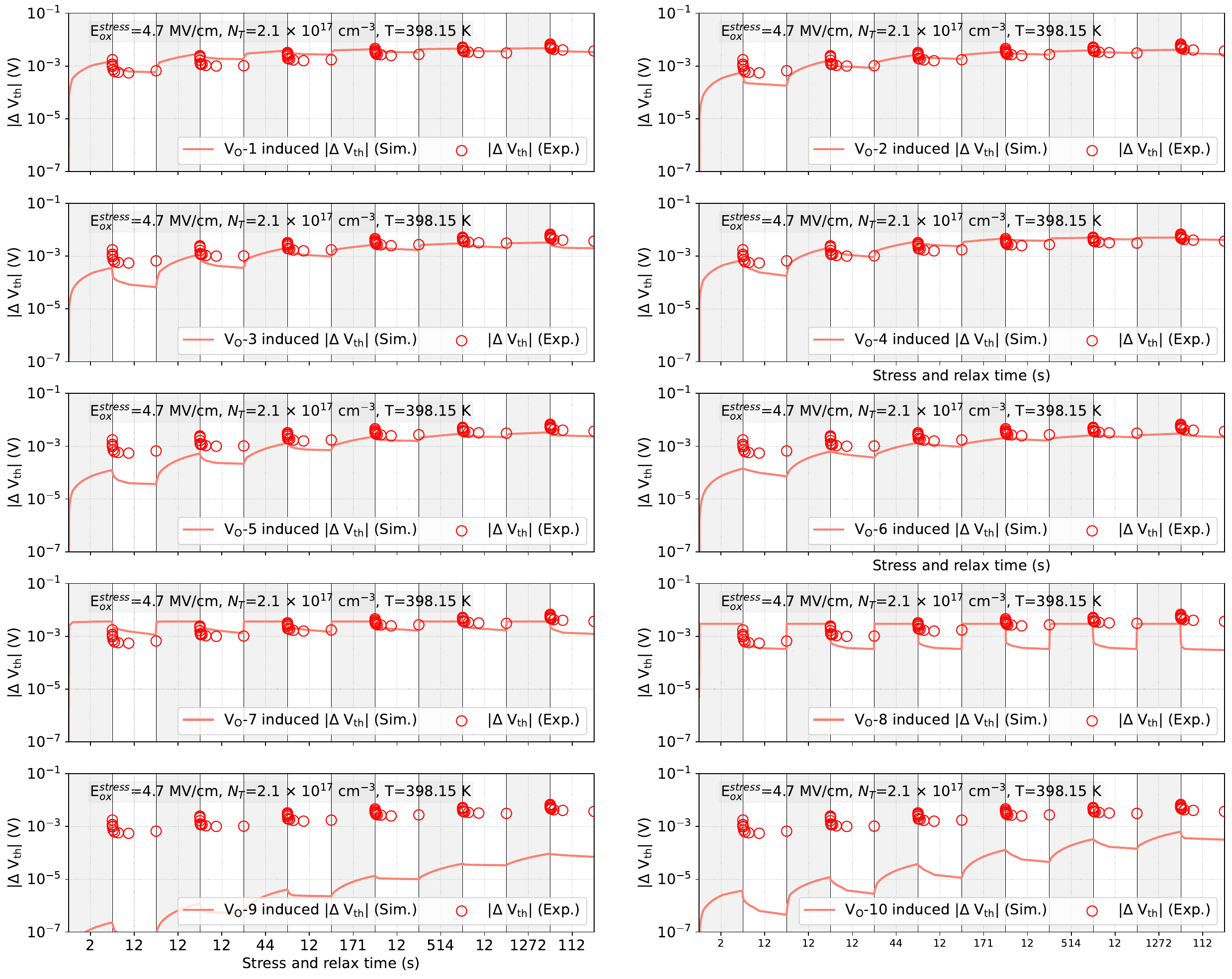
        }

        \caption{\red{Threshold voltage shift ($\Delta V_{\mathrm{th}}$) induced by V$_\mathrm{O}$ at 10 selected oxygen sites in a-SiO$_2$ during stress and recovery cycles. All simulations are performed using RASP. Each panel corresponds to a distinct oxygen site (V$_\mathrm{O}$-1 to V$_\mathrm{O}$-10). Solid red lines denote simulated $|\Delta V_{\mathrm{th}}|$ using RASP, and open red circles denote experimental data.~\cite{Comphy_Rzepa_201849, Comphy3_Dominic_2023}}}
        \label{Fig:dVth_single_site}
\end{figure*}

To quantitatively evaluate the impact of V$_{\mathrm{O}}$ defects formed at these sites on the threshold voltage shift ($\Delta V_{\mathrm{th}}$) of Si/a-SiO$_{\mathrm{2}}$ MOSFET, we performed dynamic simulations of a one-dimensional a-Si/SiO$_{\mathrm{2}}$ MOS device using RASP and compared the results with experimental measurements~\cite{Comphy_Rzepa_201849, Comphy3_Dominic_2023}. The simulations considered V$_{\mathrm{O}}$ formed at ten oxygen sites. The device structure, defect parameters, and stress/recovery condition parameters used in the simulations are listed in Table~I in Supplementary Materials. The simulation results are shown in Fig.~\ref{Fig:dVth_single_site}. The impact of V$_{\mathrm{O}}$ formed at different sites on device threshold voltage shift $\Delta V_{\mathrm{th}}$ varies significantly. Based on their different effects on device threshold voltage shift during the stress/recovery process, they can be classified into the following three categories:

\begin{enumerate}
    \item Inactive V$_{\mathrm{O}}$ traps: Such as V$_{\mathrm{O}}$-9 and V$_{\mathrm{O}}$-10, whose (+/0) transition levels are far below Si-VBM, resulting in extremely high barriers for capturing holes from Si-VBM. Therefore, under stress conditions, their hole capture rates are negligible, and their contribution to the overall device threshold voltage shift $\Delta V_{\mathrm{th}}$ is \delete{minimal}\redcnew{negligible}.
    \item Quasi-permanent V$_{\mathrm{O}}$ traps: Such as V$_{\mathrm{O}}$-(1--6), whose transition levels are relatively shallower compared to inactive V$_{\mathrm{O}}$ traps. This characteristic results in moderate hole capture rates, but the barriers for the emission process (recovery) are high, leading to extremely slow emission rates. Therefore, these defects are the main source of long-term, slow-recovery drift in devices.
    \item Fast transient V$_{\mathrm{O}}$ traps: Such as V$_{\mathrm{O}}$-(7--8), whose (+/0) transition levels are close to Si-VBM, resulting in low barriers for both hole capture and emission processes. Their hole capture and emission rates are high, which causes them to contribute significant $\Delta V_{\mathrm{th}}$ at the early stage of stress application and exhibit rapid, nearly complete recovery during the recovery phase.
\end{enumerate}

The simulation results indicate that, due to the low symmetry of amorphous materials, even for sites with the same possible defect configurations, values of $E_{T}$, $\Delta Q$, and other parameters differ in different atomic environments. Consequently, defects formed at different sites play distinctly different roles in device reliability issues. Furthermore, this low symmetry of amorphous materials also leads to different possible defect configurations in different atomic environments (as shown in Fig.~\ref{Fig:QQP_Structure} and Fig.~\ref{Fig:Structure_046}). We need to consider the impact of this low symmetry of amorphous materials on device reliability in our simulations.

In summary, when simulating the impact of defects in amorphous materials on device reliability, we need to: (1) accurately calculate parameters such as $E_{T}$, $\Delta Q$, and CC diagrams; (2) consider the contributions of all metastable configurations and their associated pathways; \delete{(3) account for the impact of defects at different sites on device reliability due to the low symmetry of amorphous materials.}\redcnew{(3) account for the site-dependent behavior of defects arising from the low symmetry of amorphous oxides.} All of these \delete{aspects }are incorporated in the all-state model used by RASP.
	
	\redc{\subsection{$\Delta V_\mathrm{{th}}$ induced by $\mathrm{V_O}$ with all possible configurations on all O sites (All-state model)}}
	\label{all}

	\begin{figure*}[!htbp]
        \centering
        \includegraphics[width=0.6\textwidth]{
            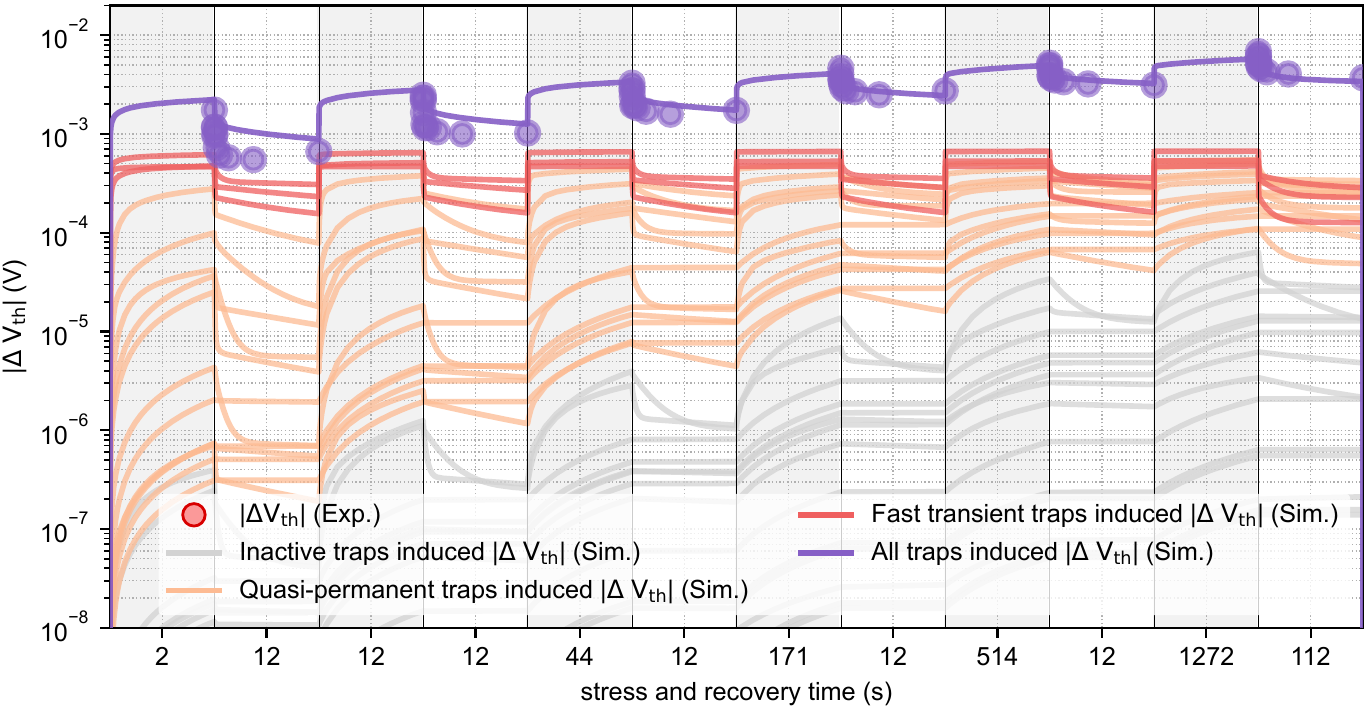
        }
       \caption{\red{Threshold voltage shift ($\Delta V_{\mathrm{th}}$) induced by V$_\mathrm{O}$ in a-SiO$_2$ during stress and recovery cycles. All simulations are performed using RASP. The simulated $|\Delta V_{\mathrm{th}}|$ is decomposed into contributions from three trap categories: fast transient traps (red curves), quasi-permanent traps (orange curves), and inactive traps (gray curves). The purple curve shows the total $|\Delta V_{\mathrm{th}}|$ contributed by all V$_\mathrm{O}$ defects. Purple circles denote the experimental $|\Delta V_{\mathrm{th}}|$ data, which are the same measurements as used for comparison in the previous figure~\cite{Comphy_Rzepa_201849, Comphy3_Dominic_2023}.}}

        \label{Fig:NBTI_SiO}
	\end{figure*}

\red{To more accurately simulate the impact of V$_{\mathrm{O}}$ in a-SiO$_{\mathrm{2}}$ on device threshold voltage shift, we used RASP with the all-state model to simulate the MOSFET threshold voltage shift caused by V$_{\mathrm{O}}$ defects in the a-SiO$_{\mathrm{2}}$ gate dielectric layer. First, we describe the sources of defect parameters used in the simulations. In our previous work~\cite{guo2024all}, we performed a global high-throughput search for all possible configurations of V$_{\mathrm{O}}^{0}$ and V$_{\mathrm{O}}^{+}$ that may form at 144 oxygen sites in a 216-atom a-SiO$_{\mathrm{2}}$ supercell, and calculated all NMP parameters at each site as input for RASP. Other parameters used in the simulations are listed in Table~I in Supplementary Materials. It should be noted that the V$_{\mathrm{O}}$ concentration used in the simulations is $8 \times 10^{17}$~cm$^{-3}$, which is consistent with experimentally measured values ($1.0 \times 10^{16}$--$1.0 \times 10^{18}$~cm$^{-3}$)~\cite{VOconcentration1, VOconcentration2, VOconcentration3, VOconcentration4}. The simulation results are shown in Fig.~\ref{Fig:NBTI_SiO}. Comparison of the device threshold voltage shift caused by V$_{\mathrm{O}}$ in a-SiO$_{\mathrm{2}}$ with experimental data~\cite{Comphy_Rzepa_201849, Comphy3_Dominic_2023} indicates that V$_{\mathrm{O}}$ in a-SiO$_{\mathrm{2}}$ is one of the possible defect sources that cannot be neglected for \delete{device }NBTI.}

\red{In previous studies, V$_{\mathrm{O}}$ was excluded as a possible source of NBTI in Si/SiO$_{\mathrm{2}}$ MOSFETs~\cite{RN76}. This conclusion was based on the four-state model, which assumes that V$_{\mathrm{O}}$ has only two stable configurations in both the neutral ($q=0$) and charged ($q=+1$) states. Specifically, the ground state of V$_{\mathrm{O}}^{0}$ is assumed to be the Si-dimer configuration and the metastable state is the back-projected configuration. After capturing a hole, the structure relaxes, with back-projected becoming the ground state and Si-dimer becoming the metastable state. This model considers two types of transitions: (1) NMP transitions involving charge state changes: V$_{\mathrm{O}}^{0}$(Si-dimer) $\leftrightarrow$ V$_{\mathrm{O}}^{+}$(Si-dimer) and V$_{\mathrm{O}}^{0}$(back-projected) $\leftrightarrow$ V$_{\mathrm{O}}^{+}$(back-projected); (2) thermal transitions involving only structural configuration changes: V$_{\mathrm{O}}^{0}$(Si-dimer) $\leftrightarrow$ V$_{\mathrm{O}}^{0}$(back-projected) and V$_{\mathrm{O}}^{+}$(Si-dimer) $\leftrightarrow$ V$_{\mathrm{O}}^{+}$(back-projected). However, simulations based on the four-state model have raised doubts about V$_{\mathrm{O}}$ as a source of NBTI defects: when V$_{\mathrm{O}}^{0}$ is in the Si-dimer ground-state configuration, its (0/+1) charge transition level is located approximately 1.55--2.47~eV below the Si valence band maximum, which is too deep. This means that even when the defect level shifts upward under negative gate bias, it remains below the \delete{channel }Fermi level, making it difficult for holes to overcome the barrier and be captured. Consequently, V$_{\mathrm{O}}$ was \delete{once }excluded as a major contributor to NBTI. In recent years, extensive research has focused on identifying new defect sources responsible for NBTI, such as hydrogen-related defects including hydrogen bridges (HB) and hydroxyl-E$^\prime$ centers (H-E$^\prime$).}
 
\red{The results in this work quantitatively demonstrate the critical role of V$_\text{O}$ in NBTI degradation through accurate simulation. This also confirms the necessity of considering defect configuration diversity (i.e., the all-state model) for accurately predicting long-term device aging behavior. Furthermore, comparing the simulation results with experimental measurements reveals that during the first two stress/recovery cycles, the recovery of $\Delta V_\text{th}$ caused by V$_{\mathrm{O}}$ is relatively slow. This suggests that other types of fast transient traps also contribute to NBTI in Si/SiO$_{\mathrm{2}}$ devices, such as hydrogen-related defects including hydrogen bridges (HB) and hydroxyl-E$^\prime$ centers (H-E$^\prime$)~\cite{RN12} that have been studied in recent years.}

\

\redc{\section{CONCLUSION}}
\label{sec:conclusion}

\redc{In this work, we develop the Reliability Ab initio Simulation Package (RASP) to address the critical challenge of accurately simulating defect-induced reliability degradation in MOSFETs. RASP implements the all-state model, which systematically considers all possible defect configurations in amorphous gate dielectrics and all nonradiative multiphonon (NMP) and thermal transition pathways among them. The package consists of four integrated modules: (i) the Device Electrostatics Module, which calculates band diagrams, electrostatic potential distributions, and carrier concentrations; (ii) the Transition Rate Module, which computes NMP transition rates, carrier tunneling rates, and thermal transition rates; (iii) the Defect Occupation Module, which models defect kinetics as a continuous-time Markov chain (CTMC) with generator matrix $\mathbf{A}(V_\mathrm{G},T)$ and solves the master equations to obtain steady-state and time-dependent probabilities of defects in each state; and (iv) the Device Reliability Module, which evaluates threshold voltage shifts through either weakly-coupled (LEVEL 1) or strongly-coupled (LEVEL 2) simulation schemes. Notably, the all-state model can be simplified to the two-state or four-state model when only the corresponding configurations and pathways are considered, both of which are supported in RASP.}

\redc{RASP introduces several methodological innovations that advance defect-based reliability simulation:} 

\redc{(i) \textit{Rapid calculation of NMP transition rate.} We employ a Fourier transform method combined with two-dimensional interpolation of the lineshape function over the ($\Delta E$, $\Delta Q$) parameter space. This approach enables rapid computation while maintaining high accuracy (10,000 defects in 87 ms). }

\redc{(ii) \textit{Efficient computation of time-dependent defect occupation probabilities under arbitrary operation.} We leverage graph-structural properties of the transition network and fundamental properties of CTMC generator matrices to determine the occupation probabilities of all defect states under arbitrary operating conditions at any given time. This enables device-level simulations to incorporate transitions among an arbitrary number of defect configurations and to quantitatively evaluate their contributions to reliability degradation. To achieve fast computation at the device simulation level, we further implement parallelized solution of the master equations, enabling efficient processing of 10,000 defects in 1.5 s.}


\redc{(iii) \textit{Site-resolved simulation of defect-induced reliability.} Conventional defect-centric models characterize the low symmetry of amorphous materials through statistical distributions of defect parameters. In contrast, RASP first evaluates the impact of defects formed at each inequivalent site on device reliability (e.g., threshold voltage shift), considering all possible configurations and transition pathways at that site. Then, RASP combines the contributions from all sites to determine the total impact on device reliability (for threshold voltage shift, this is obtained by summing the contributions from each site). This site-by-site approach eliminates the need for empirical parameter distributions.}

Using RASP to simulate threshold voltage shifts in Si/SiO$_\text{2}$ MOSFETs induced by V$_{\mathrm{O}}$, we find that V$_{\mathrm{O}}$ defects at different oxygen sites can be classified into three categories based on their contributions to $\Delta V_{\mathrm{th}}$: (i) \textit{inactive traps}, with (+/0) transition levels far below Si-VBM, exhibiting negligible hole capture rates and minimal contribution to $\Delta V_{\mathrm{th}}$; (ii) \textit{quasi-permanent traps}, with relatively shallower transition levels, serving as the main source of long-term, slow-recovery degradation; and (iii) \textit{fast transient traps}, with transition levels close to Si-VBM, exhibiting high capture and emission rates, significant $\Delta V_{\mathrm{th}}$ contribution during early stress, and rapid recovery. Taking V$_{\mathrm{O}}$ at all oxygen sites into account, we demonstrate that V$_{\mathrm{O}}$ is a non-negligible source of negative bias temperature instability (NBTI). This conclusion is contrary to previous studies based on the four-state model, \redcnew{highlighting the importance of accounting for all possible defect configurations when evaluating their impact on MOSFET reliability.}

\redc{Furthermore, other defects in amorphous gate dielectrics, such as hydrogen bridges (HB) and hydroxyl-E$^\prime$ centers (H-E$^\prime$)~\cite{RN12}, are also expected to exhibit similarly complex structural characteristics as V$_{\mathrm{O}}$\redcnew{, i.e., with various configurations and transition pathways.} Therefore, the all-state model is needed \redcnew{for quantitatively predicting the impact of these defects on device reliability. RASP can be used as a universal tool for such studies in device reliability physics as well as the design of high-reliablility devices.}}

{\color{black} \section*{\label{sec:level1}Data availability}

    The software and data will be made available on request.
    }
\begin{acknowledgments}
    This work was supported by National Natural Science Foundation of China (12334005,
    12188101 and 12404089).
\end{acknowledgments}

\bibliography{apssamp}

\end{document}